%% file: main.tex
\documentclass[a4paper,fleqn,usenatbib]{mnras}
\usepackage{mystyle}

\input{./title_page}

\begin{document}

\newcommand{\pdd}[2]{\frac{\partial #1}{\partial #2}}

\label{firstpage}
\pagerange{\pageref{firstpage}--\pageref{lastpage}}
\maketitle

\input{./abstract}

\input{./introduction}
\input{./methods}
\input{./results}
\input{./discussion}
\input{./conclusion}

\input{./acknowledgements}

\bibliographystyle{mnras}
\bibliography{export}

%\appendix
%\input{./appendix}

\bsp
\label{lastpage}
\end{document}

%% file: title_page.tex
\title[Formation of early-type galaxies]{Forming early-type galaxies without AGN feedback: a combination of merger-driven outflows and inefficient star formation}

\author[M. Kretschmer \& R. Teyssier]{
Michael Kretschmer$^{1}$\thanks{E-mail: michael.kretschmer@physik.uzh.ch},
Romain Teyssier$^{1}$
\\
$^{1}$Institute for Computational Science, University of Zurich, Winterthurerstrasse 190, CH-8057 Zurich, Switzerland\\
}

\date{Accepted XXX. Received YYY; in original form ZZZ}

\pubyear{2019}

%% file: abstract.tex
\begin{abstract}
Regulating the available gas mass inside galaxies proceeds through a delicate balance between inflows and outflows, but also through the internal depletion of gas due to star formation. At the same time, stellar feedback is the internal engine that powers the strong outflows. Since star formation and stellar feedback are both small scale phenomena, we need a realistic and predictive subgrid model for both. 
We describe the implementation of supernova momentum feedback and star formation based on the turbulence of the gas in the RAMSES code.
For star formation, we adopt the so-called multi-freefall model.
The resulting star formation efficiencies can be significantly smaller or bigger than the traditionally chosen value of $1\%$.
We apply these new numerical models to a prototype cosmological simulation of a massive halo 
that features a major merger which results in the formation of an early-type galaxy without using AGN feedback.
We find that the feedback model provides the first order mechanism for regulating the stellar and baryonic content in our simulated galaxy.
At high redshift, the merger event pushes gas to large densities and large turbulent velocity dispersions, such that efficiencies come close to $10\%$, 
resulting in large $\mathrm{SFR}$. We find small molecular gas depletion time during the starburst, in perfect agreement with observations. 
Furthermore, at late times, the galaxy becomes quiescent with efficiencies significantly smaller than $1\%$, resulting in small $\mathrm{SFR}$ and long molecular gas depletion time.
\end{abstract}


%% file: introduction.tex
\section{Introduction}

Although galaxy formation still remains one of the most important unsolved problems in astrophysics,
we have made spectacular progresses in the past decade. 
It became apparent that galaxy formation has to be very inefficient at forming stars, with a peak 
value around 20\% of the available baryons for Milky Way sized haloes, down to much smaller efficiencies at lower and larger masses
\citep{1992MNRAS.258P..14P,1998ApJ...503..518F,2000MNRAS.311..793B,2001MNRAS.326..255C,2001MNRAS.326.1228B,2006RPPh...69.3101B}.
One of the key observables that triggered 
our recent advances comes from the abundance matching technique \citep{2010MNRAS.404.1111G,2010ApJ...710..903M,2013ApJ...770...57B},
from which we know relatively precisely the stellar mass of the central galaxy as a function of the parent halo mass.

The inefficiency of galaxy formation led our community to discover the importance of galactic winds to regulate star formation \citep{1986ApJ...303...39D,2006MNRAS.373.1265O}, a theoretical idea corroborated
(some might say incepted) by observations of strong outflows at high redshift \citep{2003ApJ...588...65S,2010ApJ...717..289S}.
In parallel, it was recognised that cosmological infall of fresh gas, in the form of cold streams, was also an important player in setting up the star formation rate
in high redshift galaxies \citep{2005MNRAS.363....2K,2008MNRAS.390.1326O,2009Natur.457..451D}.
Galaxy mergers are now believed to play a minor role in the history of gas accretion, while major mergers (mass ratios close to 1:1)
trigger the extreme and rare star bursts we see nearby, as well as in the more distant universe \citep{2017ARA&A..55...59N}.

Regulating the available gas mass inside galaxies proceeds through a delicate balance between inflows and outflows \citep{2010MNRAS.406.2325O},
but also through the internal depletion of gas due to star formation. In the same time, stellar feedback is the internal
engine that powers the strong outflows, and since star formation and stellar feedback are both small scale phenomena, we need a realistic
and predictive subgrid model for both. Although it has been argued that strongly star forming galaxies have their star formation rate regulated by feedback \citep{2014MNRAS.445..581H},
the transition to quenched galaxies, and their extremely low star formation rate still remains a puzzle. Most galaxies in the low redshift universe are quiescent, even disk
galaxies like the Milky Way. It is therefore of paramount importance to use the right model of star formation to get the correct 
star formation efficiency in this quiescent, quasi-quenched regime. Active Galactic Nuclei (AGN) feedback is invoked to explain the origin of massive, red and dead elliptical galaxies 
\citep{1998A&A...331L...1S,2005MNRAS.361..776S,2011MNRAS.414..195T,2011ASL.....4..204B}. 
At intermediate masses, quiescent, spheroidal and dispersion dominated galaxies can be modelled without relying on AGN feedback, but on a 
combination of strong early feedback leading to gas removal and suppression of gas cooling due to a hotter halo temperature \citep{2007ApJ...658..710N, 2010ApJ...709..218F,2014MNRAS.444.3357N}.
Even in these extremely quiescent galaxies, star formation is still active, but in a quenched, very inefficient mode \citep{2011MNRAS.415...32S,2011MNRAS.415...61S}. 

Quantitatively, the efficiency of star formation is characterised by the {\it gas depletion time},
defined as the ratio of the star formation rate to the available gas mass. 
Observationally, the depletion time can be estimated using the so-called global Kennicutt-Schmidt (KS) relation
\citep{1998ApJ...498..541K}. It turns out that this gas depletion time is very long in nearby galaxies, 
typical around 1~Gyr, with even longer depletion time around 3~Gyr for more massive quiescent galaxies \citep{2011MNRAS.415...61S}.

More information can be obtained using the {\it resolved} KS relation \citep{2007ApJ...671..333K}. One can compute 
the local value of the depletion time in patches of roughly 1~kpc in size. This more local KS relation demonstrates
than the local star formation rate {\it surface density} scales with the local gas surface density, with a power law slope
around 1.5 \citep{2007ApJ...671..333K}. This leads to the simple theoretical idea that the star formation rate
can be modelled locally using a so-called Schmidt law
\begin{equation}
\dot{\rho}_\star=\epsilon_{\rm ff} \frac{\rho}{t_\mathrm{ff}}~~~{\rm with}~~~t_\mathrm{ff}=\sqrt{\frac{3\pi}{32 G \rho}}
\end{equation}
where $\dot{\rho}_\star$ is the local star formation rate density, $\epsilon_{\rm ff}$ is a {\it fixed} star formation efficiency
per freefall time, 
$\rho$ is the local gas density and 
$t_\mathrm{ff}$ is the local gas freefall time. This simple model naturally explains the observed power law.
It is usually used only if the gas density lies above a fixed threshold corresponding to star forming gas. 
This fixed efficiency can be chosen to match the observed relation in nearby resolved galaxies \citep[e.g.][]{2006MNRAS.373.1074S,2018MNRAS.475.3283C}.
The value usually adopted in cosmological simulations of galaxy formation lies around $ \epsilon \sim 0.01$ \citep[see e.g.][]{2011mnras.410.1391A}. 
Note that the star formation efficiency in these models is by construction uniform throughout the galaxy.

On smaller scales, close to individual Giant Molecular Clouds (GMC) of size between 1 and 10~pc, the situation is not so simple.
Although the average star formation efficiency is again between 1 and 2\% \citep{2007ApJ...654..304K}, it seems that $\epsilon_{\rm ff}$ can vary significantly 
from cloud to cloud \citep{2011ApJ...729..133M,2010ApJ...724..687L}. Although this effect can be explained by a spurious
systematic effect in the observational protocol \citep{2011ApJ...727L..12F}, it could also have a physical origin,
related to different internal properties in the star forming molecular clouds. 

Current theories of star formation,
based on self-gravitating supersonic turbulence, favour a scenario where star formation is the result of a turbulent cascade.
In this framework, one can derive a local star formation efficiency by integrating over the log-normal distribution of 
the turbulent gas density \citep{2005ApJ...630..250K,2011ApJ...743L..29H,2012ApJ...761..156F}. In particular, the so-called {\it multi-freefall model} 
of star formation \citep[see e.g.][]{2012ApJ...761..156F} give promising results, 
with $\epsilon_{\rm ff}$ depending on two important cloud structural parameters, namely the cloud mean density 
and the cloud mean Mach number (see below for more details). 

Recently, cosmological simulations of galaxy formation started exploring this new approach to model star formation.
For example, \cite{2018MNRAS.480..800H} used the GIZMO code together with a method that requires the gas to be self-gravitating, 
self-shielded and Jeans-unstable to form stars, very close in spirit to the multi-freefall approach.
Using the RAMSES code, several models based on the multi-freefall approach were presented by \cite{2015IAUGA..2257403P} and 
\cite{2017MNRAS.470..224T,2018MNRAS.478.5607T} with different implementation details.
In an isolated galaxy, \cite{2016ApJ...826..200S} and \cite{2018MNRAS.474.2884L} explored a similar model based on the local star formation efficiency models of \cite{2012ApJ...759L..27P} and \cite{2011ApJ...730...40P} respectively.
All these different studies revealed that the local star formation efficiency in the simulated galaxies is very inhomogeneous,
and varies widely, both in time and space, similar to what we observe at small scales in molecular clouds.
Additionally, on the larger galactic scales, either globally or using 1~kpc patches, the observed KS relations were also successfully reproduced.

Reproducing the KS relation on large scale could have nothing to do with the details of star formation 
at small scales. Star formation on large scales could be entirely self-regulated by stellar feedback, 
as advocated by \cite{2010ApJ...721..975O} and \cite{2014MNRAS.445..581H}. 
In a seminal albeit recent paper, \cite{2018ApJ...861....4S} have shown that once the star formation efficiency at small scale is large enough (typically 
larger than 1\%), the global depletion time converges to a fixed value set by stellar feedback only (therefore independent of $\epsilon_{\rm ff}$). 

However, for lower star formation
efficiencies (typically smaller than 1\%), this is not true anymore and the global depletion time becomes longer and longer, 
inversely proportional to $\epsilon_{\rm ff}$.
The transition between these two regimes occurs around 1\%. 
This critical value depends on the strength of the adopted feedback model: A weaker stellar feedback model leads to shorter 
gas ejection phases and a shorter depletion time scale. This gives a larger critical value for $\epsilon_{\rm ff}$. 
It is therefore of primary importance to use the correct local star formation model
{\it and} the correct feedback model to get the right transition between the quiescent regime and the feedback dominated regime.

Numerical recipe for modelling stellar feedback have seen tremendous progresses over the past decade.
Although several stellar evolution processes are believed to inject thermal and kinetic energy into the
surrounding interstellar medium (ISM) of a star or a star cluster \citep{2011MNRAS.417..950H,2013ApJ...770...25A},
it is now believed that Type II supernovae explosion is the dominant mechanism for stellar feedback. The 
challenge is then to resolve the cooling radius that marks the transition from the energy conserving phase of 
the remnant to its momentum conserving phase. This allows to correctly describe the hot X-ray emitting gas in the ISM
and to accumulate enough momentum to get the right amount of terminal momentum in the final phase. 

The first numerical implementations
were based on suppressing cooling artificially and increasing the cooling radius to the spatial resolution of the simulation
\citep[see e.g.][and references therein]{2013MNRAS.429.3068T}.
Although this model is qualitatively successful in modelling strong supernovae feedback, 
it overestimates the hot gas fraction and the terminal momentum of the explosion. It was later proposed
to directly inject the correct thermal energy and terminal momentum into neighbouring cells \citep{2011MNRAS.417..950H,2013ApJ...770...25A}.
This method is now widely adopted, as it seems to capture the right amount of kinetic energy, but as shown by \cite{2019MNRAS.483.3363H}, 
it does not capture the hot phase properly.
These different methods were recently compared and cross-calibrated in \cite{2017MNRAS.466...11R}, who concluded that no recipe is truly satisfactory,
but direct injection of the correct terminal momentum seems to be the least bad method of all.

In this paper, we use the RAMSES code, for which several competing supernovae feedback implementations have been developed over the past years.
\cite{2013ApJ...770...25A} injected directly the correct supernovae terminal momentum, 
using a as non-thermal energy variable and modifying the Riemann solver accordingly. 
\cite{2014MNRAS.444.2837R} explored the effects of radiative feedback, within the framework of the \cite{2013MNRAS.429.3068T} delayed cooling model. 
Finally, \cite{2014ApJ...788..121K,2015MNRAS.451.2900K} implemented a mechanical feedback scheme that directly injects the correct terminal momentum in the surrounding cells, 
based on the earlier model of \cite{2008A&A...477...79D}.

In this paper, we combine these various new developments, both for star formation with a varying efficiency
and for supernovae momentum feedback, combining the best of each past implementation in a novel and unique way,
that we believe is superior to what has been done before in the RAMSES code.
We apply this new implementation of subgrid galaxy formation physics to a prototype cosmological simulation of a massive halo 
that features a major merger, and study specifically the interesting case of an early-type galaxy.
We are particularly interested in the effect of our local star formation recipe on the global galaxy properties.
For this, we explore two different local star formation models: a classical Schmidt law with uniform efficiency 
and the multi-freefall model proposed by \cite{2012ApJ...761..156F}.
We finally discuss why this new model can explain the properties of quenched galaxies,
as a consequence of strong feedback combined with a very inefficient local star formation.

%% file: methods.tex
\section{Numerical methods}

We use the RAMSES Adaptive Mesh Refinement (AMR) code to model the dynamical evolution of both dark matter and baryonic fluids
in a cosmological context \citep{2002A&A...385..337T} .
AMR allows to divide space into cubical cells that can be adaptively refined to increase the spatial resolution locally
according to some adopted refinement criterion. In this work, we use the so-called {\it quasi-Lagrangian} strategy,
for which cells are refined when the dark matter or baryonic mass exceeds a given threshold value, that corresponds to
the mass resolution adopted in the initial conditions (see below for details). For simulations that include baryonic physics, one must impose a maximum level of 
resolution (or a minimal cell size) because of limited computational resources. Current state-of-the-art cosmological galaxy formation
simulations reach a spatial resolution between tens to hundreds of parsecs, barely resolving the vertical thickness of the gaseous disks,
and still missing the scale of the largest molecular clouds in the Galaxy. Only very recently do we see a few examples of ``cloud resolving''
galaxy simulations in the cosmological context \citep{2017MNRAS.472.2356M,2019MNRAS.490.4447W,2020MNRAS.491.1656A}, while isolated, idealised simulations are already capable of resolving molecular clouds
down to sub-pc scales \citep{2015MNRAS.446.2038R,2017MNRAS.470L..39F,2019MNRAS.483.3363H}. None of these simulations are however able to resolve the truly star-forming scale, namely 
individual subsonic (or mildly transonic) molecular cores below 0.1~pc. 

It is therefore mandatory to augment these numerical models with a subgrid description of the physics of the ISM. Subgrid models of galaxy formation
have been introduced more than 20 years ago to model star formation in cosmological simulations
using the previously discussed Schmidt law \citep{1992ApJ...399..331C,1992ApJ...399L.109K}. 
As explained in the introduction, our community has now realised that we need to go beyond this 
simple recipe to model the physics of star formation. Although a complete theory of star formation is still to be found, we have made tremendous 
progresses in the past couple of decades \citep{2007ARA&A..45..565M}, the key physical ingredient being supersonic turbulence \citep{2004RvMP...76..125M}. 
Modern subgrid models of star formation and feedback all feature, with different levels of sophistication, a model for supersonic turbulence. 
We first describe our approach of the problem. We then explain how this subgrid turbulence model is coupled to a physically motivated recipe for star formation and stellar feedback.
 
\subsection{Subgrid model for turbulence}
\label{sec:subgrid_sgs}

Turbulence can be modelled numerically using the Navier-Stokes equations while resolving the microscopic viscous diffusion scale.
This approach, called Direct Numerical Simulations (DNS) in engineering applications, is not realistic for galaxy formation,
as under normal ISM conditions the viscous scales are many orders of magnitude smaller than any achievable resolution. 
In the 60's, \cite{1963MWRv...91...99S} proposed a model (later called Large Eddy Simulations or LES) for which the small scales are averaged over to
define the mean flow. The fluid equations are modified through this averaging procedure, introducing the turbulent pressure and several turbulent diffusion terms.
The LES approach allows to model turbulent flows without resolving the dissipative scale, at the expense of designing a Sub-Grid Scale (SGS) model to describe turbulent
effects at the macroscopic scale. Note that numerical dissipation already naturally provides an implicit SGS model for turbulence, sometimes referred to as {\it implicit LES models.}

In astrophysics, LES and SGS models have been introduced by \cite{2006A&A...450..265S} and \cite{2006A&A...450..283S} to study Type Ia supernovae turbulent flame combustion. 
The subgrid turbulence is used to compute the turbulent flame speed properly, accounting for unresolved eddies in the reacting flow. 
Later, the formalism was extended by \cite{2011A&A...528A.106S} to supersonic turbulent flows and applied to the physics of the ISM. 
Finally, in the context of galaxy formation, \cite{2016ApJ...826..200S} introduced a very similar SGS model, coupled to a subgrid star formation model, in the spirit of what we present here. 

The difficulty with supersonic turbulence is to model both velocity and density fluctuations. The mass density is decomposed into
the average, large scale density $\overline{\rho}$, defined as the volume-averaged density field, smoothed at the scale of the cell $\Delta x$
and the fluctuation $\rho'$, so that the subgrid density writes as $\rho = \overline{\rho} + \rho'$. The average velocity 
field and average temperature are however defined using a mass-weighted average (also called the Favre average) with
\begin{equation}
\widetilde{v} = \frac{\overline{\rho v}}{\overline{\rho}}~~~{\rm and}~~~\widetilde{T} = \frac{\overline{\rho T}}{\overline{\rho}},
~~~{\rm with}~~~v = \widetilde{v}+v''~~~{\rm and}~~~T=\widetilde{T}+T'',
\end{equation}
where the fluctuations are now defined relative to the Favre average with a double prime. The turbulent kinetic energy is finally defined as
\begin{equation}
K_T=\frac{1}{2} \overline{\rho v''^2} = \frac{1}{2} \overline{\rho} \sigma_{\rm 3D}^2,
\end{equation}
where we introduce the turbulent three-dimensional velocity dispersion $\sigma_{\rm 3D}=\sqrt{3}\sigma_{\rm 1D}$.
Using these definitions, it is possible to derive new fluid equations for the mean flow variables $\overline{\rho}$, $\overline{v}$ and $\overline{T}$,
as well as a new equation for the turbulent kinetic energy. For details on the derivation and the complete form of these new equations, 
we refer to \cite{2011A&A...528A.106S} and \cite{2014nmat.book.....S}. Note that in this paper, contrary to the strategy adopted in \cite{2016ApJ...826..200S},
we do not consider the modified form of the Euler equations. We follow the philosophy of \cite{2006A&A...450..265S} and rely on numerical diffusion
to provide an implicit LES model, without adding extra diffusion to an already too diffusive numerical approach. We only consider the extra equation
on the turbulent kinetic energy, that writes \citep{2014nmat.book.....S,2016ApJ...826..200S}
\begin{equation}
\pdd{}{t}K_{\rm T} + \pdd{}{x_j} \left( K_{\rm T} \widetilde{v_j} \right) + P_{\rm T} \pdd{\widetilde{v_j}}{x_j} = C_{\rm T} - D_{\rm T},
\end{equation}
where the turbulent pressure $P_{\rm T}=2/3 K_{\rm T} = \overline{\rho} \sigma_{\rm 1D}^2$.
The {\it creation term} $C_{\rm T}$ is prescribed in the eddy viscosity model 
(also called mixing length theory in different contexts) using the mean flow viscous stress as
\begin{equation}
C_{\rm T} = 2 \mu_{\rm T} \sum_{ij}
\left[ \frac{1}{2}\left( \pdd{\widetilde{v_i}}{x_j}+\pdd{\widetilde{v_j}}{x_i} \right)
- \frac{1}{3} \left( \nabla \cdot \widetilde{\bf v} \right) \delta_{ij}\right]^2 = \frac{1}{2} \mu_{\rm T} \left| {\widetilde S}_{ij}\right|^2.
\end{equation}
The {\it destruction term} $D_{\rm T}$ represents the dissipation of turbulence in the subgrid turbulent cascade down to viscous scales.
It is modelled as
\begin{equation}
D_{\rm T} = \frac{K_{\rm T}}{\tau_{\rm diss}}.
\end{equation}
The two important parameters in the SGS turbulence model are thus the turbulent viscosity parameter $\mu_{\rm T}$
and the turbulent dissipation time scale $\tau_{\rm diss}$. They are both related to the cell size (also called
in different contexts the smoothing length or the mixing length) by
\begin{equation}
\mu_{\rm T} = \overline{\rho} \Delta x \sigma_{\rm 1D}~~~{\rm and}~~~\tau_{\rm diss} = \frac{\Delta x}{\sigma_{\rm 1D}}.
\end{equation}
Note the strong analogy with the Chapman-Enskog theory for viscous rarefied gases, except that individual colliding particles
are replaced here by viscous eddies.

Other recent implementations of subgrid models for star formation use an instantaneous estimate of the turbulent velocity dispersion 
\citep{2015IAUGA..2257403P,2018MNRAS.480..800H,2017MNRAS.470..224T,2018MNRAS.478.5607T}. This approach can be interpreted here as the stationary limit of 
our turbulent kinetic energy equation, for which
\begin{equation}
C_{\rm T} = D_{\rm T} \Longrightarrow \sigma_{\rm 1D} = \frac{\Delta x}{\sqrt{3}} \sqrt{\left| {\widetilde S}_{ij}\right|^2}.
\end{equation}
Solving the full turbulent kinetic energy equation allows to account for advection and work of turbulent pressure, as well as 
non-equilibrium dissipation of turbulent kinetic energy. There is however an important caveat in the SGS approach: The
turbulent creation term is modelling injection of turbulent kinetic energy through shear flows. In the absence of gravity,
shear flows are indeed always unstable and turbulent, owing to the famous Kelvin-Helmholtz instability. In presence of gravity, however, this not necessarily true anymore,
as gravity might stabilise the flow, owing for non-convective conditions. On the other hand, self-gravity in a marginally stable disk
might be an additional source of turbulence that we do not consider explicitly in our model.
Our SGS model might overestimate the amount of turbulence, especially in an equilibrium, centrifugally supported disk,
or underestimate it in a gravitationally unstable disk.
This is why we will explore alternative models of creation of turbulence in the Results section.

\subsection{Subgrid model for star formation}
\label{sec:subgrid_sf}

Once we know the turbulent kinetic energy in each cell, we can prolong the turbulent spectrum to smaller,
unresolved scales $\ell < \Delta x$ using Burgers turbulence spectrum, for which
\begin{equation}
\sigma(\ell) = \sigma_{\rm 1D} \left( \frac{\ell}{\Delta x} \right)^{1/2}.
\end{equation}
A critical unresolved scale is the sonic length, defined by $\sigma(\ell_s) = c_s$ and given be
$\ell_s = \Delta x / \mathcal{M}^2$,
where the cell Mach number is defined by $\mathcal{M} = \sigma_{\rm 1D}/c_s$.
These scales correspond to the subsonic (or mildly transonic) molecular cores,
in which stars form. Below the sonic scale, density fluctuations become very weak. 
Each molecular core can thus be seen as a quasi-homogeneous region of space that 
will eventually collapse and form a star. On scales larger than the sonic length, however,
density fluctuations are very large.
Following \cite{2012ApJ...761..156F}, but adapting here slightly their methodology, we assume that
gas density distribution in this supersonic turbulent medium follows a log-normal PDF, with
\begin{equation}
p(s)=\frac{1}{\sqrt{2\pi \sigma_s^2}}\exp {\frac{(s-\overline{s})^2}{2\sigma_s^2}}
\label{eq:pdf-distr} 
\end{equation}
with the logarithmic density $s=\ln (\rho / \overline{\rho})$ where $\rho$ is the local density and $\overline{\rho}$ the mean density of the cell.
The distribution is normalised such that 
\begin{equation}
\int_{-\infty}^{+\infty} p(s){\rm d}s = 1 
~~~{\rm and}~~~
\int _{-\infty}^{+\infty} \rho p(s){\rm d}s = \overline{\rho}.
\end{equation}
The mean logarithmic density $\overline{s}=-1/2 \sigma_s^2$ is related to the standard deviation $\sigma_s$ which can be fitted, using non-magnetised,
isothermal turbulence simulations \citep{2011ApJ...730...40P}, by
\begin{equation}
\sigma_s^2 = \ln \left( 1 + b^2 \mathcal{M}^2\right),
\end{equation}
where $b$ is a parameter related to the exact nature of the turbulence forcing (solenoidal or compressive).

Following the model of \cite{2005ApJ...630..250K} for star formation and assuming that these homogeneous cores
are spherically symmetric with diameter $\ell_s$, we can compute their virial parameter as
\begin{equation}
\alpha_s = \frac{2 E_\mathrm{kin}}{|E_\mathrm{grav}|} = \frac{15}{\pi} \frac{c_s^2+\sigma(\ell_s)^2}{G \rho \ell_s^2} .
\end{equation}
A molecular core will collapse and form stars if $\alpha_s < 1$. This can be translated into a critical density threshold for star formation
\begin{equation}
\rho > \rho_{\rm crit} = \frac{15}{\pi} \frac{2 c_s^2 \mathcal{M}^4}{G \Delta x^2} = \alpha_{\rm vir} \overline{\rho} \frac{2 \mathcal{M}^4}{1+\mathcal{M}^2},
\end{equation}
where we defined the virial parameter of the whole cell as
\begin{equation}
\alpha_{\rm vir} = \frac{15}{\pi} \frac{c_s^2+\sigma_{\rm 1D}^2}{G \overline{\rho} \Delta x^2} = \frac{15}{\pi} \frac{c_s^2 }{G \overline{\rho} \Delta x^2} (1+\mathcal{M}^2).
\label{eq:def-alpha-vir}
\end{equation}
Finally, we derive the lognormal critical density for star formation as
\begin{equation}
s_\mathrm{crit} = \ln \left[ \alpha_\mathrm{vir} \frac{2 \mathcal{M}^4}{1+\mathcal{M}^2}\right].
\end{equation}

\begin{figure}
\includegraphics[width=\columnwidth]{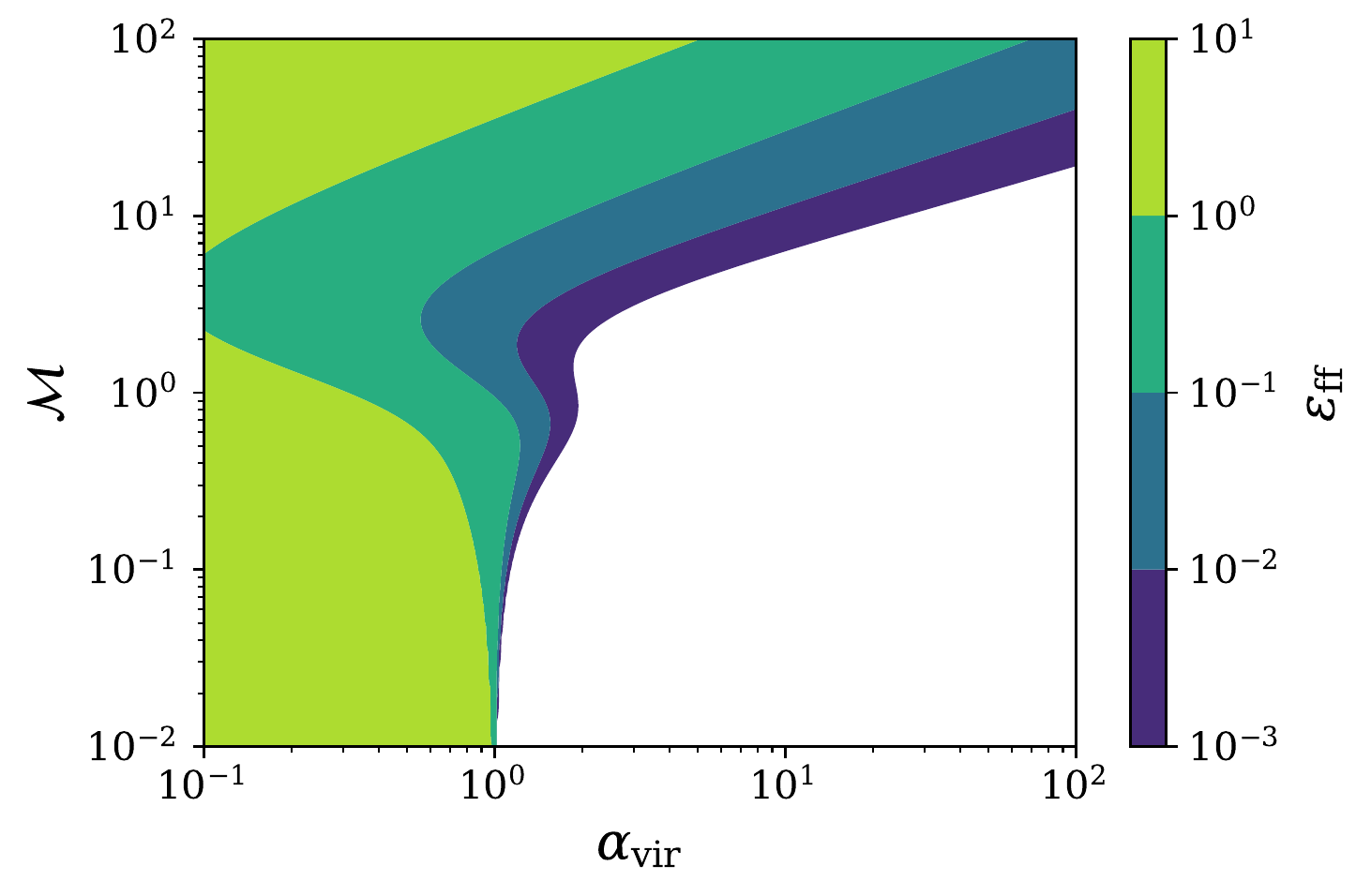}
\caption{Star formation efficiency per freefall time as a function of the two parameters $\alpha_\mathrm{vir}$ and $\mathcal{M}$ in our modified multi-freefall model. 
For $\mathcal{M} \lesssim 1$, SF is only efficient if $\alpha_\mathrm{vir} \lesssim 1$ corresponding to the gravitational collapse of the whole computational cell. 
For higher values of $\mathcal{M}$, at fixed $\alpha_\mathrm{vir}$, the efficiency increases because of turbulent compression inside the cell at small, unresolved scales.}
\label{fig:sf_model}
\end{figure}

In the {\it multi-freefall models} of \cite{2012ApJ...761..156F}, the local star formation rate is expressed as
\begin{equation}
\dot \rho_* = \int_{s_\mathrm{crit}}^{\infty}\frac{\rho}{t_\mathrm{ff}(\rho)} p(s) \diff s = \epsilon_{\rm ff} \frac{\overline{\rho}}{t_{\rm ff}(\overline{\rho})}.
\end{equation}
This formulation just states that each fluid element that satisfies the gravitational instability criterion collapses
in one free-fall time and converts all its mass into (one or several) stars. From this, we can deduce the local star formation efficiency
\begin{equation}
\begin{split}
\epsilon_\mathrm{ff} &= \int_{s_\mathrm{crit}}^{\infty}\frac{t_\mathrm{ff}(\overline{\rho})}{t_\mathrm{ff}(\rho)}\frac{\rho}{\overline{\rho}} p(s) \diff s \\
&= \frac{1}{2} \exp \left( \frac{3}{8} \sigma_s^2\right)\left[ 1 + \mathrm{erf}\left( \frac{\sigma_s^2-s_\mathrm{crit}}{\sqrt{2 \sigma_s^2}}\right) \right].
\label{eq:sfr-ff}
\end{split}
\end{equation}
According to \cite{2012ApJ...761..156F}, the model breaks down for $\mathcal{M} \leq 2$. 
Indeed, for $\mathcal{M} = 1$, the sonic length becomes equal to the cell size. 
In the simulations we present here, we have most of the time $\mathcal{M} \geq 10$,
but not all the time. Many cells can have subsonic turbulence, especially in the hot gas phase.
We need to provide a model that is valid also for low Mach numbers.
We thus modify the collapse criterion for $\mathcal{M} < 1$, requiring now the whole cell to be gravitationally unstable.
\begin{equation}
\alpha_s = \frac{15}{\pi} \frac{c_s^2+\sigma_{\rm 1D}^2}{G \rho \Delta x^2} < 1.
\end{equation}
Note that we didn't use the mean cell density but the local density $\rho$ in the previous equation, so now the density PDF 
plays the role of a probability for the entire cell to have a certain density, while $\overline{\rho}$ is just the expectancy. We finally have
\begin{equation}
s_\mathrm{crit} = \ln \left[ \alpha_\mathrm{vir} \right]~~~{\rm for}~~~\mathcal{M} \leq 1.
\end{equation}
In order to account for both regime $\mathcal{M} \leq 1$ and $\mathcal{M} \geq 1$ in a single equation, we propose here to modify the original \cite{2012ApJ...761..156F} model 
by combining them into one single critical density
\begin{equation}
s_\mathrm{crit} = \ln \left[ \alpha_\mathrm{vir} \left( 1 + \frac{2 \mathcal{M}^4}{1+\mathcal{M}^2} \right)\right],
\end{equation}
that we use in conjunction with Equation~\ref{eq:sfr-ff}. \autoref{fig:sf_model} shows the resulting star-formation efficiency per freefall time $\epsilon_\mathrm{ff}$ as a function of the two parameters $\alpha_\mathrm{vir}$ and $\mathcal{M}$. 
For low Mach numbers, the efficiency for star formation goes sharply from 0 if $\alpha_{\rm vir}>1$ to 1 otherwise. For an isothermal gas, the corresponding recipe is just a simple density threshold, allowing star formation
only when the Jeans length is not resolved anymore by the mesh. This method has been used for a long time in galaxy formation simulations \citep[see e.g.][]{2013MNRAS.429.3068T}. For larger Mach numbers, however, the efficiency iso-contours become wider.
Gas with $\alpha_\mathrm{vir}>1$ is allowed to collapse, owing to the large density fluctuations caused by supersonic turbulence, although the efficiency can be significantly lower than 100\%. Typical conditions for star forming regions
in the Milky Way are $\alpha_\mathrm{vir} \simeq 3-10$ and $\mathcal{M} \simeq 10-20$, and correspond to $\epsilon_{\rm ff} \simeq 0.01-0.02$, in good agreement with observations \citep[see discussion in ][]{2016ApJ...826..200S}.
Traditional models of galaxy formation often adopt a similar value for their fixed $\epsilon_\mathrm{ff}=0.01$ \citep{2006A&A...445....1R,2011mnras.410.1391A}.

We now discuss some of the main caveats of the present approach. First, we have assumed a very simple spherical geometry for the collapsing transonic cores.
This is not true in reality, where filamentary structures are very often seen in star forming regions \citep{2014prpl.conf...27A}. This effect could be parametrised through a multiplicative fudge factor 
in front of $\alpha_{\rm vir}$ in the collapse criteria. 

Second, in the multi-freefall formulation, we have assumed that within each transonic core the efficiency of star formation is 100\%, and no gas is left after stars are born.
In our view, stellar feedback, as implemented in the next section, is responsible for terminating star formation. It is however possible to parametrise the efficiency of star formation per transonic cores using another 
multiplicative fudge factor in front of the integral in Equation~\ref{eq:sfr-ff}. 

Third, we have also considered collapsing fluid elements at the sonic scales, but larger, supersonic fluid elements could also collapse
at a smaller density, further fragmenting into individual protostars. Such a model has been proposed for example by \cite{2011ApJ...743L..29H} and \cite{2012MNRAS.423.2037H}.

Fourth, our lognormal model for the density fluctuations is only strictly speaking applicable to isothermal supersonic turbulence. At the relatively large scales considered here with $\Delta x \simeq$ 100~pc,
the ISM is far from strictly isothermal. This can affect the density distribution in a non-trivial way \citep{2008ApJ...680.1083R}. Having these caveats in mind, we nevertheless believe that any modification of the 
proposed model will not affect our results strongly, and certainly not at a qualitative level.

\subsection{Subgrid model for stellar feedback}

Stellar evolution is probably the most extreme subgrid aspect in galaxy formation. 
In principle, we need to follow the evolution of the internal structure of the stars, 
describe their evolution throughout their main sequence, until they die and explode,
at least for the most massive ones. Although it is reasonable to assume that stellar interiors are fully decoupled
from galactic scales, this is not true for the final explosive stages of stellar evolution. 

\begin{figure}
\includegraphics[width=\columnwidth]{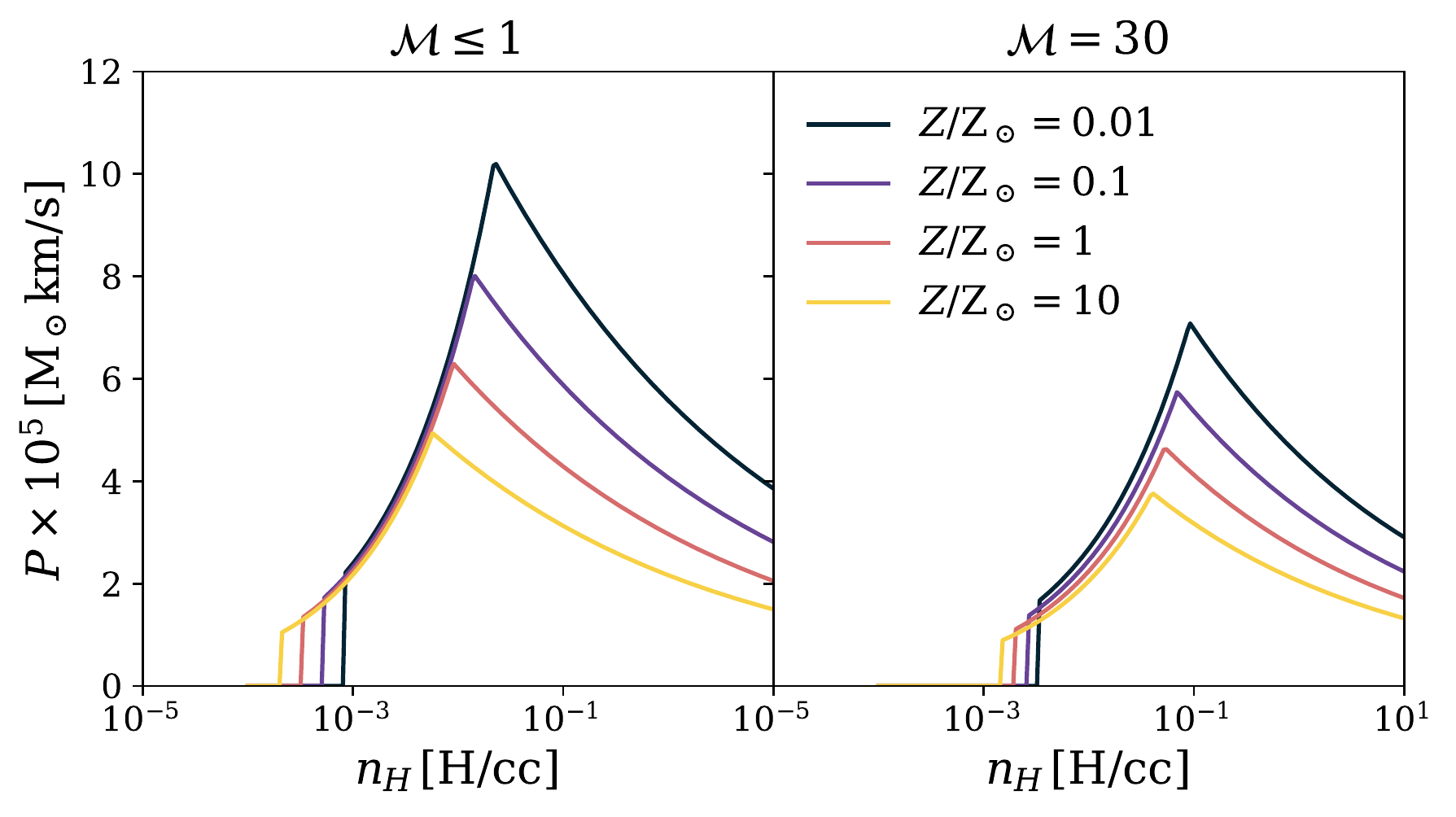}
\caption{Injected scalar momentum for a single supernova explosion as a function of density. A resolution of $\Delta x_\mathrm{min}= 150 \mathrm{pc}$ is assumed here. 
Different colours show the injected momentum for various metallicities. 
For low density cells, we don't inject momentum. For intermediate density cells, the cooling radius is resolved by at least one cell, but less than four cells. 
We compute the injected momentum using the Sedov solution. For high density cells, for which the cooling radius is not even resolved by one cell, we inject the terminal momentum.
The left plot shows the injected momentum for a homogeneous medium, while the right plot corresponds to an inhomogeneous medium with $\mathcal{M}=30$ \citep[from][]{2015MNRAS.450..504M}.}
\label{fig:mom-feedback}
\end{figure}

Since our stellar particles have a typical mass of $\sim 10^5 \mathrm{M}_\odot$, we have to account for individual supernovae using a subgrid model.
For this, we assume that supernovae explosions within this single stellar population are uniformly distributed between $t_\mathrm{start} = 3$ and $t_\mathrm{end}=20$ Myrs. 
At each simulation time step, we compute the supernova rate as
\begin{equation}
\dot{N}= \frac{\eta_\mathrm{SN} M_\mathrm{ini}}{ M_\mathrm{SN} (t_\mathrm{end} - t_\mathrm{start})},
\end{equation}
where $\eta_\mathrm{SN}=0.2$ is the mass fraction of the population that explodes in supernovae, $M_\mathrm{ini}$ is the initial mass of the stellar particle and $M_\mathrm{SN}=10 \mathrm{M}_\odot$ is the typical mass of a single progenitor.
Multiplying by the time step $\Delta t$, we obtain the expected (average) number of supernovae $\langle N \rangle = \dot{N} \Delta t$. The actual number of exploding supernova for a given stellar particle during the time step is drawn from a 
Poisson distribution \citep[see][for a similar technique]{2018MNRAS.477.1578H}. This allows us to have supernovae explosions as discrete events 
realistically distributed in time, independent on the adopted mass resolution.

The other challenge of modelling supernovae explosions comes from the requirement of resolving the energy-conserving Sedov phase of the remnant.
This phase is absolutely crucial as it is responsible for the momentum build-up that will later efficiently accelerate the surrounding gas.
The scale that marks the transition from energy-conserving to momentum-conserving is the cooling radius $R_\mathrm{cool}$.

If the cooling radius is unresolved by the grid, which usually is the case for high gas densities, the thermal energy will be spuriously radiated away, without having the chance to create enough momentum.
Therefore, to properly simulate the effects of supernova on the gas, we inject momentum additionally to the thermal injection if the cooling radius is not resolved. This technique,
usually referred to as kinetic feedback or momentum feedback, has progressively emerged as one of the best subgrid models for supernovae explosions \citep{2014MNRAS.445..581H,2014ApJ...788..121K,2017ApJ...834...25K,2016ApJ...824...79A,2017MNRAS.466...11R,2018MNRAS.478..302S,2018MNRAS.480..800H}.

On the other hand, if the cooling radius is resolved by the grid, which typically happens for low density gas, the supernova energy is directly injected in the form of thermal energy. 
The energy-conserving blast wave that will follow will then properly accelerate the gas around it and deliver the correct amount of momentum.

The terminal momentum of the momentum-conserving blast wave and the corresponding cooling radius can be computed analytically for homogeneous background densities and simple cooling functions.
For more realistic cases, including explosions within a supersonic turbulent medium, we can use high-resolution simulations of individual explosions, such as in \cite{2015MNRAS.450..504M}. 
Using these numerical models, we use for the cooling radius
\begin{equation}
R_\mathrm{cool} = 3.0\, \mathrm{pc}\left(\frac{Z}{Z_\odot}\right)^{-0.082} \left( \frac{n_\mathrm{H}}{100\,\mathrm{cm}^{-3}}\right)^{-0.42},
\end{equation}
where $n_\mathrm{H}$ is the gas density of the cell where the star is exploding.
If $R_\mathrm{cool}$ is unresolved by at least one grid cell, we inject for each individual supernova the terminal momentum $P_\mathrm{SN}$
\begin{equation}
P_\mathrm{SN} = 1.42 \times 10^5 \mathrm{km\,s^{-1}M_\odot} \left(\frac{Z}{Z_\odot}\right)^{-0.137}\left(\frac{n_\mathrm{H}}{100\,\mathrm{cm}^{-3}}\right)^{-0.16}
\end{equation}
Note that in the previous two equations, we use a metallicity floor at $0.01Z_\odot$ to model the effect of primordial cooling for a pristine gas. 
The total amount of injected scalar momentum from each star particle into its surrounding cell is
\begin{equation}
P = P_\mathrm{SN} N_\mathrm{SN} \min \left( 1,\left(\frac{\Delta x_{\rm min}}{R_\mathrm{cool}}\right)^{3/2}\right)
\label{eq:tot-mom}
\end{equation}
where we use the Sedov solution when $\Delta x_{\rm min} < R_{\rm cool} < 4 \Delta x_{\rm min}$. 
Here $N_\mathrm{SN}$ is the number of supernovae for the current time step and for each star particle, while $\Delta x_{\rm min}$ is the minimum AMR cell size.
\autoref{fig:mom-feedback} shows the dependence of the injected momentum on the gas density for one supernova and for an adopted resolution of $\Delta x_\mathrm{min}= 150 \mathrm{pc}$.
From \cite{2015MNRAS.450..504M}, we designed two models shown in \autoref{fig:mom-feedback}. The first one is valid for cells with a low Mach number, while the second one applies for high turbulent Mach number.
In this paper, we use the low Mach number solution, but ideally, one should use the Mach number as an additional parameter. This model is unfortunately unavailable at the present time.

Once we know the injected scalar momentum per star particle, we need to deposit isotropically this momentum into the grid.
For this, we use novel numerical techniques, inspired by the work of \cite{2013ApJ...770...25A} and \cite{2014MNRAS.445..581H}. 

First, we deposit the individual particle momentum onto the grid using the cloud-in-cell interpolation technique. We obtain the grid scalar momentum density $p = P/\Delta x^3$ and convert it into a momentum flux density using 
\begin{equation}
Q = p \Delta x/\Delta t = P/\Delta x^2 /\Delta t
\end{equation}
Note that $Q$ is analogous to a pressure. This momentum flux is split equally among the 6 cell faces, defining a supernovae dynamical pressure noted $P_\star=Q/6$. This pressure only accounts for the momentum flux due to the
supernovae explosions. It is directly added to the thermal pressure in the Riemann solver. This direct injection of momentum through the Riemann solver was first explored by \cite{2013ApJ...770...25A}, using non-thermal energy and pressure variables
that are available in the RAMSES code. This method deposits the right amount of momentum to the gas, but also affects the thermal energy with undesirable effects (spurious heating or cooling of the gas). 

We follow here a different approach by adding only this new pressure term in the momentum equation 
\begin{equation}
\pdd{}{t}\left( \rho v_i \right)+ \pdd{}{x_j} \left( \rho v_i v_j \right) + \pdd{}{x_i} \left( P+P_\star \right) = -\rho \pdd{\phi}{x_i}
\end{equation}
and remove its corresponding $p{\rm d}V$ work from the energy equation, 
so that the internal energy equation remains unaffected by this momentum injection
\begin{equation}
\pdd{E}{t}+ \pdd{}{x_j} \left( E+P+P_\star \right) v_j = P_\star \pdd{v_j}{x_j} - \rho \pdd{\phi}{x_j} v_j
\end{equation}
This approach, a slight modification of the original method from \cite{2013ApJ...770...25A}, will deliver the proper momentum flux to neighbouring cells, together with consistent mass and energy fluxes.
We also modify the time step stability condition, increasing the wave speed in the Courant condition to $|v| + c_s + p/\rho$, 
where $p$ is here the scalar momentum density defined above.

Our new method is more efficient than the traditional recipe for kinetic feedback in the RAMSES code, as implemented in \cite{2008A&A...477...79D} and \cite{2015MNRAS.451.2900K}.
These models inject the right solution at the relevant time of the deposition. In the limit of very poor resolution, they effectively inject the terminal momentum along with thermal energy.
However, in these earlier attempts, the momentum is deposited on the grid using an explicit spherical velocity profile spread over a sphere of 4 cells in radius. This is usually done every coarse step and is quite costly. This is incompatible with our objective of resolving discrete supernovae with possibly a very high frequency of explosions.

On the other hand, depositing the momentum in only the 6 direct neighbours of the supernovae triggers spurious grid-aligned effects, especially if multiple supernovae explode in the same cell \citep{2018MNRAS.477.1578H}. 
To avoid this, we choose the location of each individual explosion at random among the 8 cloud-in-cell neighbours from the star particle, suppressing visible grid alignment effects.

Although momentum feedback has proven quite successful in delivering the right amount of kinetic energy to the surrounding gas, a significant improvement over previous delayed cooling recipe, it still suffers 
from the caveat of not resolving properly the hot diffuse phase filling the cavity bounded by the momentum-conserving dense gas shell. This could affect the wind properties. 
\cite{2014MNRAS.443.1173H} have shown recently in the context of a dwarf galaxy that the main properties of the outflow are indeed captured by the momentum injection scheme,
but properly describing the multiphase temperature structure of the wind requires to resolve the supernovae cooling radius.

Additionally, we model HII regions around young stars using a simple recipe for which we maintain the gas temperature (only in the cell where the star sits) at $10^4$~K for 20~Myr, until the last supernovae explodes.
Thermal feedback from each supernova is modelled by injecting $E_\mathrm{SN} = 10^{51} \mathrm{erg}$ in the parent cell. Furthermore, we assume that each supernova ejects $1 \mathrm{M}_\odot$ of metal into the ISM.

%% file: results.tex
\section{Numerical experiments}

We now report on the numerical experiment we have performed to test these new implementations of the key subgrid models
for galaxy formation. For this, we model the cosmological evolution of a single dark matter halo, whose mass is comparable to the 
Milky Way mass, but whose accretion history is much more violent. In particular, we choose a halo with one major merger at relatively low redshift
(expansion factor $a>0.5$). This will give us the opportunity to study how star formation and feedback behave in extreme cases. As it turned out, this particular halo
features a strong starburst at $a=0.65$, followed by an extremely quiescent state, that we identify as a typical example of the population of 
early-type galaxies with long gas depletion time scales \citep{2011MNRAS.415...61S}. It is worth stressing 
that in our model we don't use AGN feedback, so that any quenching mechanism here must be a consequence of our adopted stellar feedback and/or star formation models.

\subsection{Simulation setup}

We base our analysis on a cosmological zoom-in simulation that we performed with the AMR code RAMSES \citep{2002A&A...385..337T}.
We first ran a reference pure N-body simulation with $512^3$ particles in a periodic box of size 25~$h^{-1}$Mpc. 
First, we have selected all halos with masses in the range of $5 \times 10^{11} \leq M_\mathrm{vir}/\mathrm{M}_\odot \leq 1.5 \times 10^{12}$, where the virial mass is defined as the mass contained in a sphere of radius $R_\mathrm{vir}$ that encompasses an over-density of $200$ times the critical density.
Secondly, we applied an isolation criterion at $z=0$ where halos are excluded if there is another halo within $5 R_\mathrm{vir}$ with a mass of $10\%$ of the target halo.
Finally, we looked at the mass-accretion histories of the available halos and selected one halo which experienced its last merger at $a=0.65$. This halo has a mass of $M_\mathrm{vir}=6.5 \times 10^{11}$M$_\odot$ at $z=0$ and was used for our analysis.

We then generated new initial conditions around this halo with the MUSIC code \citep{2011MNRAS.415.2101H}, with an initial hierarchy of concentric grids from $\ell_{\rm min}=7$, corresponding to a coarse grid resolution of $128^3$ covering the entire periodic box, to $\ell_{\rm max,ini}=10$, corresponding to an effective initial resolution of $1024^3$. 
This gives us a dark matter particle mass of $m_{\rm dm}=1.7 \times 10^{6}$M$_\odot$ and a baryonic initial mass resolution of $m_{\rm dm}=2.6 \times 10^{5}$M$_\odot$.

We then performed 
several simulations including gas and galaxy formation physics. The maximum resolution was set to $\ell_{\rm max}=18$ at $z=0$, while refinement levels were 
progressively released to enforce a constant physical resolution of $\Delta x_{\rm min}=100 h^{-1}$pc. The adopted refinement criterion is the traditional quasi-Lagrangian 
approach, namely cells are individually refined when more than 8 dark matter particles are present or when the baryonic mass (gas and stars) exceed $8\times m_{\rm bar}$. 
Only the Lagrangian volume corresponding to twice the final virial radius of the halo was refined, the rest of the box being kept at a fixed, coarser resolution to provide
the proper tidal field. Our zoom simulation corresponds to a halo slightly smaller than the Milky Way, but it features two successive major mergers, one at $a \simeq 0.35$
with a stellar mass ratio of 1:1 and a stellar mass of each galaxy of $M_* \simeq 1.5 \times 10^{9}$M$_\odot$, and one later at $a \simeq 0.65$ with a stellar mas ratio of 1:1 too, and a larger galaxy stellar mass of $M_* \simeq 5 \times 10^{9}$M$_\odot$. This scenario turned out to be an ideal one to study extreme regime of 
star formation and analyse the impact of our star formation and feedback recipe. 

Gas cooling and heating is implemented using equilibrium chemistry for Hydrogen and Helium \citep{1996ApJS..105...19K}, together with the metallicity dependent cooling function 
for metals from \cite{1993ApJS...88..253S}. Additionally, self-shielded UV-heating of the gas is taken into account, where the self-shielding density is assumed to be 
$n_\mathrm{H}=0.01 \mathrm{H/cc}$ \citep{2010ApJ...724..244A}. We adopt the model of \cite{1996ApJ...461...20H} for the UV background.

We ran three different simulations with different models for the galaxy formation physics. 
\begin{enumerate}
\item The first simulation was run without feedback but using our new multi-freefall star formation model. It is labelled ``no feedback'' in the figures. 
\item The second simulation was run with stellar feedback but adopting the old school constant efficiency model with $\epsilon_{\rm ff}=0.01$
and a star formation density threshold of $\rho_*=0.1$H/cc. It is labelled ``constant efficiency'' in the figures.
\item The third simulation was run with both our new recipe, namely stellar feedback and our varying $\epsilon_{\rm ff}$ model. It is labelled ``varying efficiency'' in the figures.
\end{enumerate}
Note that we could have run the second simulation with a higher density threshold, to mimic a criterion $\alpha_{\rm vir}<1$ for some fixed temperature and Mach number.
This would give results intermediate between our two cases. Note that if one adopts a high density threshold, $\rho_*=10$H/cc for example, it is impossible to form star in lower density gas, so that quenching occurs automatically in this case, in a rather ad-hoc way. 
Our multi-freefall approach, on the other hand, allows for star formation even in gas with $\alpha_{\rm vir}>1$ if the subgrid turbulence is strong enough.

We finally list our more technical RAMSES settings for these 3 simulations. We used the HLLC Riemann solver, modified to account for the supernovae dynamical pressure.
The star particle was set equal to the baryonic mass resolution $m_*=m_{\rm bar}$. We adopted the {\it MinMod} slope limiter, an important choice to ensure
the stability of the new feedback scheme. 
The adiabatic exponent of the gas was set to $\gamma=5/3$. 
We didn't use any polytropic pressure floor, as our star formation model, by construction, very quickly removes gas for which the Jeans length 
is not resolved. The gas metallicity was initialised to $Z_{\rm ini}=10^{-3}Z_\odot$ to account for early population III stars enrichment.

\begin{figure*}
\minipage{0.32\textwidth}
  \includegraphics[width=\linewidth]{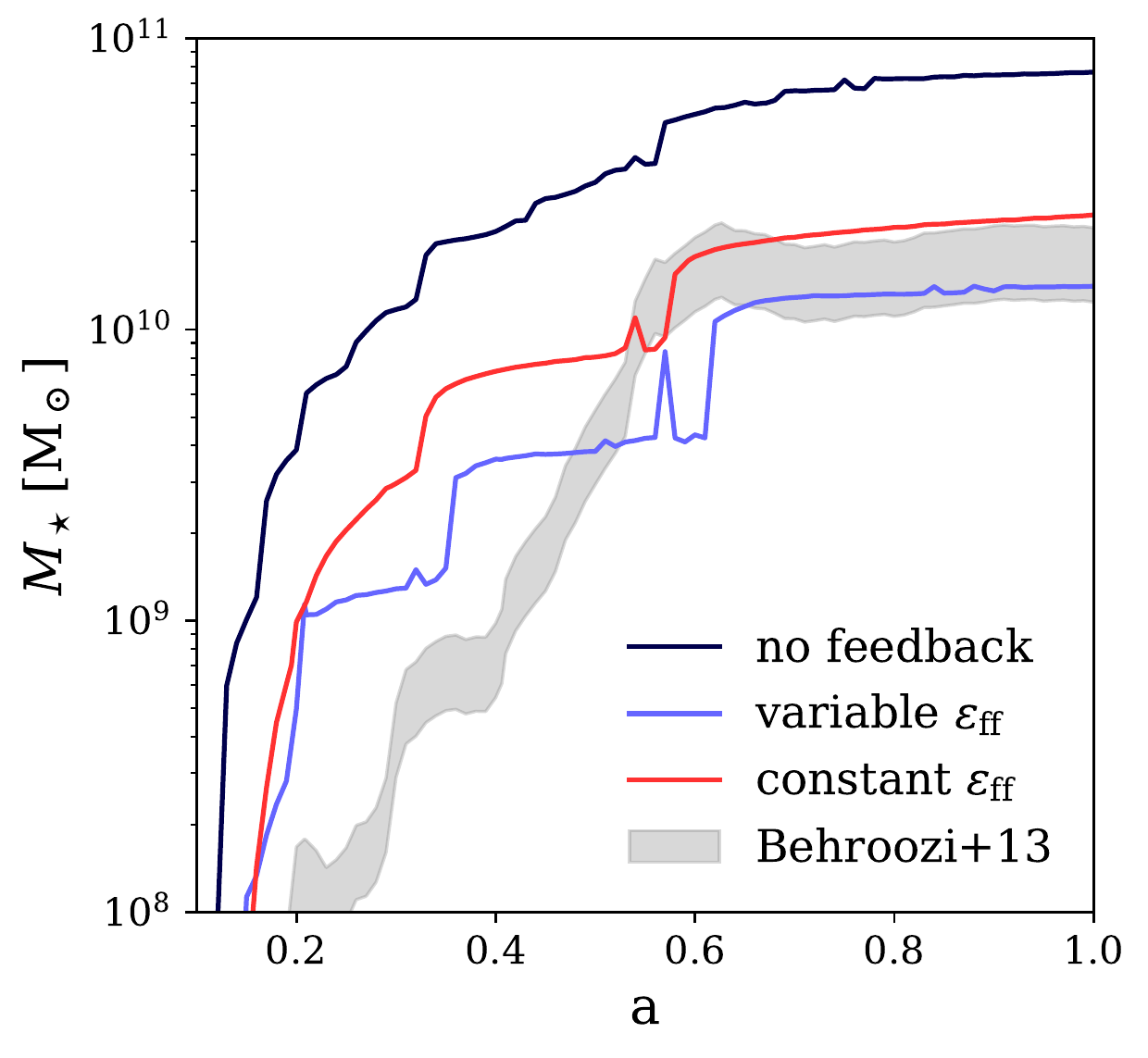}
\caption{Evolution of the stellar mass as a function of the expansion factor in our 3 simulations. The shaded area represents the prediction from abundance matching according to
\protect\cite{2013ApJ...770...57B} using the virial mass of the halo at each time.}
\label{fig:am}
\endminipage\hfill
\minipage{0.32\textwidth}
  \includegraphics[width=\linewidth]{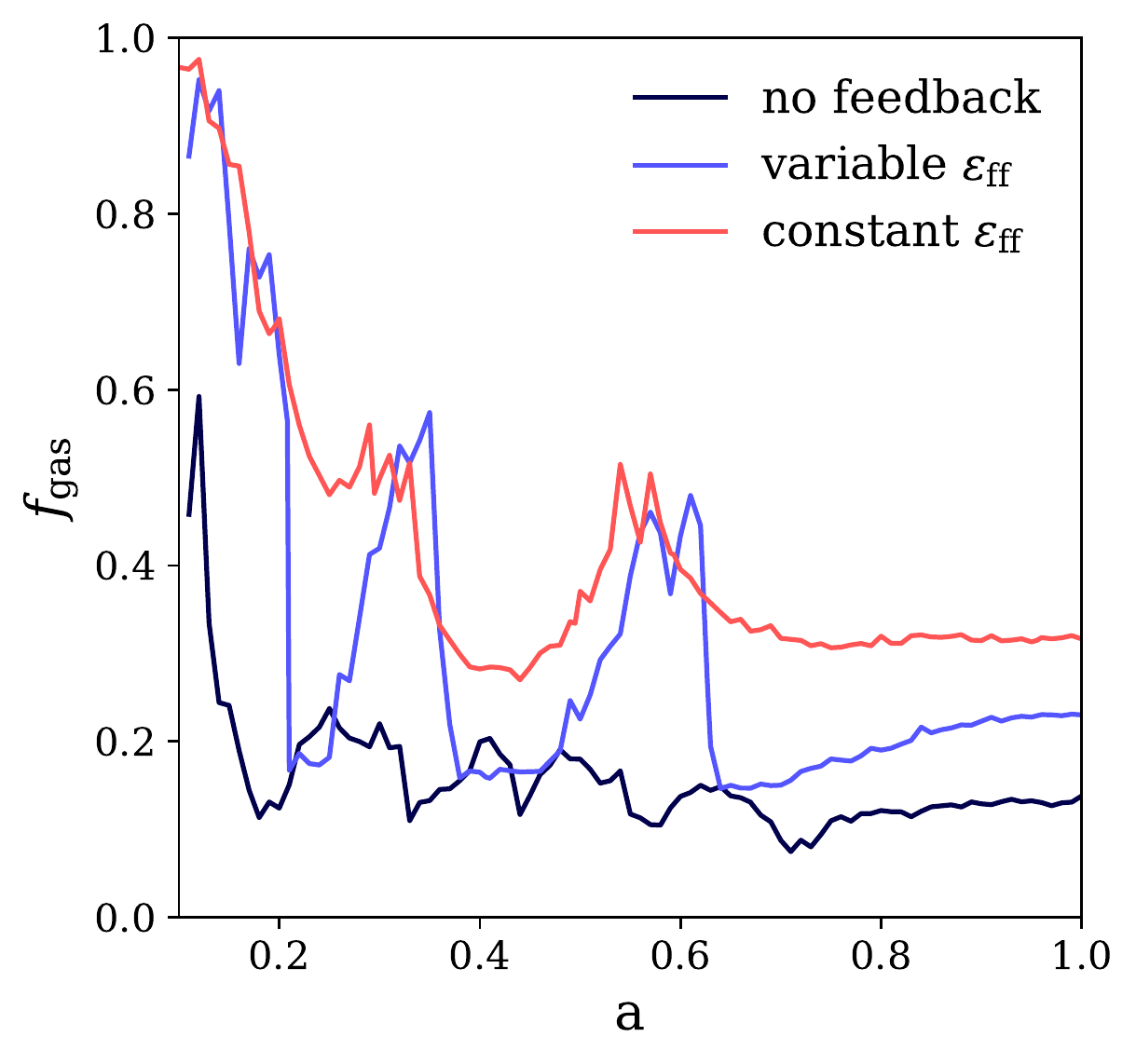}
\caption{Evolution of the gas fraction inside $0.1 R_\mathrm{vir}$ as a function of expansion factor for our 3 simulations. 
Sharp drops correspond to strong starburst-driven outflows with a large gas mass being removed.}
\label{fig:f_gas}
\endminipage\hfill
\minipage{0.32\textwidth}%
  \includegraphics[width=\linewidth]{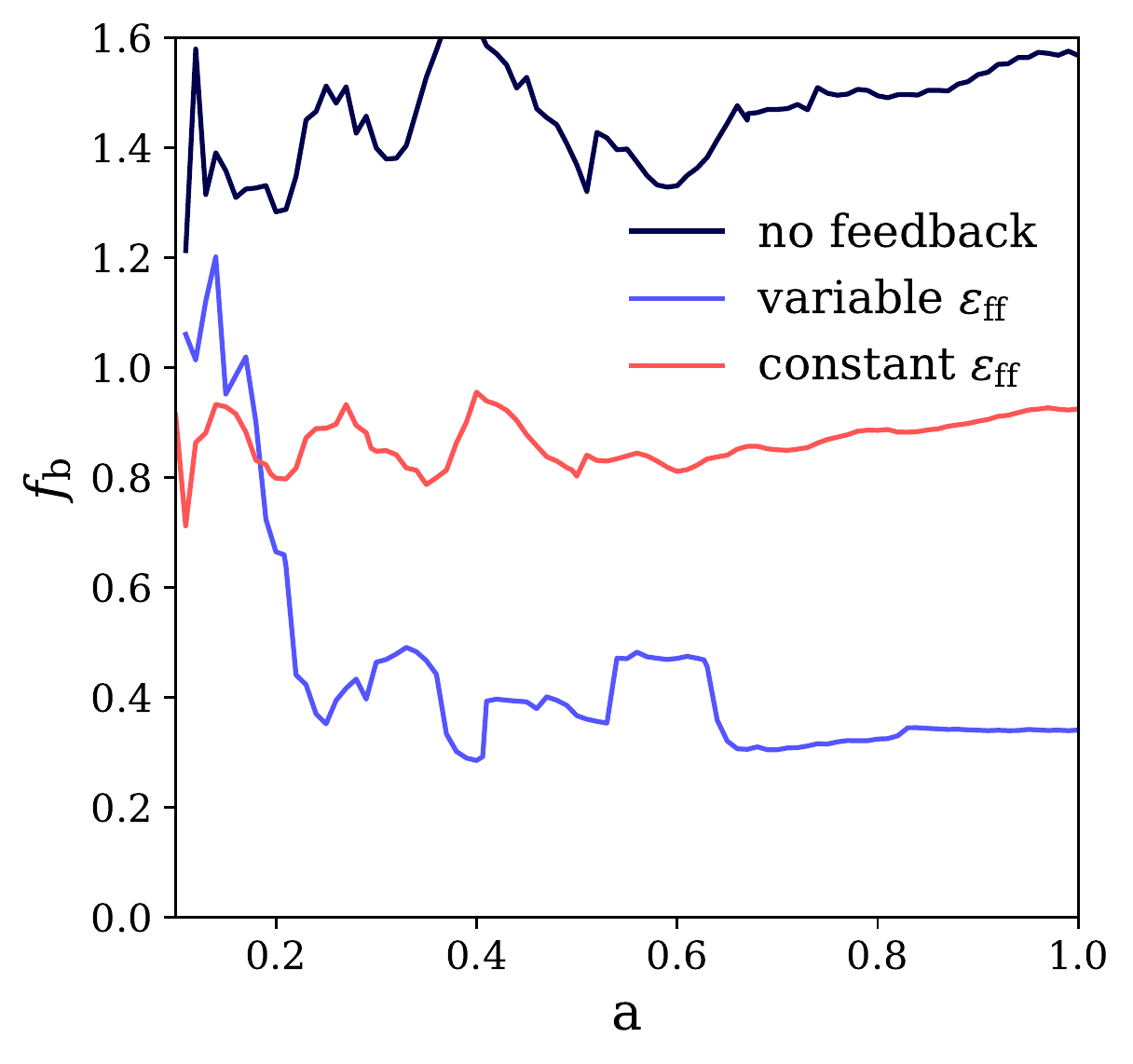}
\caption{Evolution of the baryon fraction inside the virial radius as a function of expansion factor for our 3 simulations. 
In the no feedback run, the baryon fraction is 40\% higher than the universal value, while in the varying efficiency model, 60\% of the available baryons have been ejected
outside the virial radius. }
\label{fig:bar_frac}
\endminipage
\end{figure*}

\subsection{Effect of the feedback model}

We plot in \autoref{fig:am} the evolution of the stellar mass of the central galaxy in our simulated galaxy for the 3 different subgrid models. 
The central galaxy is defined here as everything within $0.1 \times R_{\rm vir}$ from the centre. As a validation test, we also show as a shaded area 
the expected stellar mass of the central galaxy from abundance matching \citep{2013ApJ...770...57B}, using the simulated halo mass as input at each epoch. 
The simulation without feedback completely overestimates the stellar mass (by a factor of 5), while our two models with feedback reproduce roughly the expected evolution.
We see a clear trend for the constant efficiency model to produce almost twice as many stars than the multi-freefall model. 
The two major mergers in the mass accretion history of this particular halo can be seen as two jumps in the stellar mass of the central galaxy at $a \simeq 0.35$ and $a \simeq 0.65$.

In \autoref{fig:f_gas}, we show the evolution of the gas fraction within $0.1 \times R_{\rm vir}$, defined by $f_\mathrm{gas} = M_\mathrm{gas}/(M_\mathrm{gas} + M_\star)$. 
Ironically, the lowest gas fraction is obtained for the no feedback run, as most of the gas is consumed and turned into stars. The constant efficiency model, on the other hand,
has the largest gas fraction, with $f_{\rm gas} \simeq 0.4$ at late time, while the varying efficiency model shows larger fluctuations, with peaks of gas fraction
during star bursts, and a lower value of $f_{\rm gas} \simeq 0.2$ at late time, in better agreement with observation (see Discussion section). 
Our interpretation is that stellar feedback in the varying efficiency model is stronger, owing to the higher efficiency in the star bursts mode (see next Section).
This is corroborated by \autoref{fig:bar_frac} that shows the baryon fraction with $R_{\rm vir}$. The no feedback case manages to accumulate more baryons 
in the virial radius than the universal fraction, while feedback maintains the baryon fraction below the universal value in all cases. The varying efficiency case,
however, only retained 40\% of the baryons in the halo, meaning that 60\% of the baryons are lurking outside the virial radius, probably never coming back.

\subsection{Effect of the star formation model}

\begin{figure}
\includegraphics[width=\columnwidth]{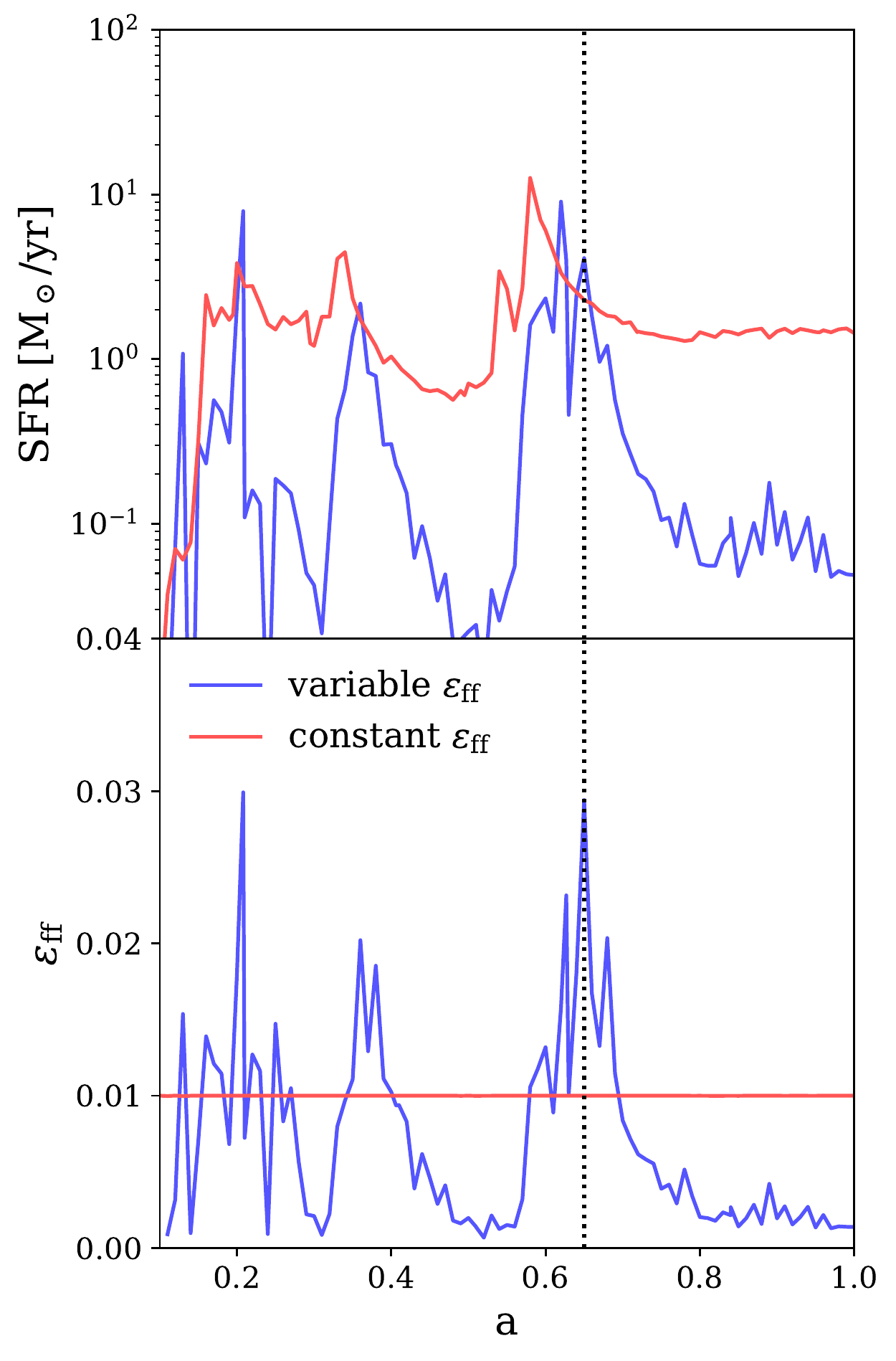}
\caption{Top panel: The star formation history of the central galaxy as a function of the expansion factor for our two star formation models. 
Bottom panel: The average star formation efficiency per free fall time as a function of the expansion factor (see the exact definition in the text). }
\label{fig:sfr_eps}
\end{figure}

Although the stellar feedback subgrid model seems to play the main role in regulating the stellar mass of the galaxy, 
the star formation subgrid model seems to have a non-negligible impact on our results.
The top panel of \autoref{fig:sfr_eps} shows the star formation rate (SFR) as a function of the scale factor for our 2 models with feedback.
It is defined as the instantaneous SFR within $0.1 \times R_{\rm vir}$ at each time.
We ignore the model without feedback in what follows because we believe it is not realistic.
We clearly see the two major mergers as two strong starbursts with the SFR peaking at 10~$M_\odot~{\rm yr}^{-1}$. 
We used a bin size of $\Delta a=0.01$ corresponding roughly to 100~Myr at low redshifts. 

In the varying efficiency case, the SFR during the starbursts is slightly larger than for the constant efficiency model. But this is after the starbursts that 
the difference between the 2 models is striking. The SFR for the varying efficiency model drops significantly down to 0.1~$M_\odot~{\rm yr}^{-1}$, while it remains
quite high, around 1~$M_\odot~{\rm yr}^{-1}$ for the constant efficiency case.

\begin{figure*}
\includegraphics[width=\textwidth]{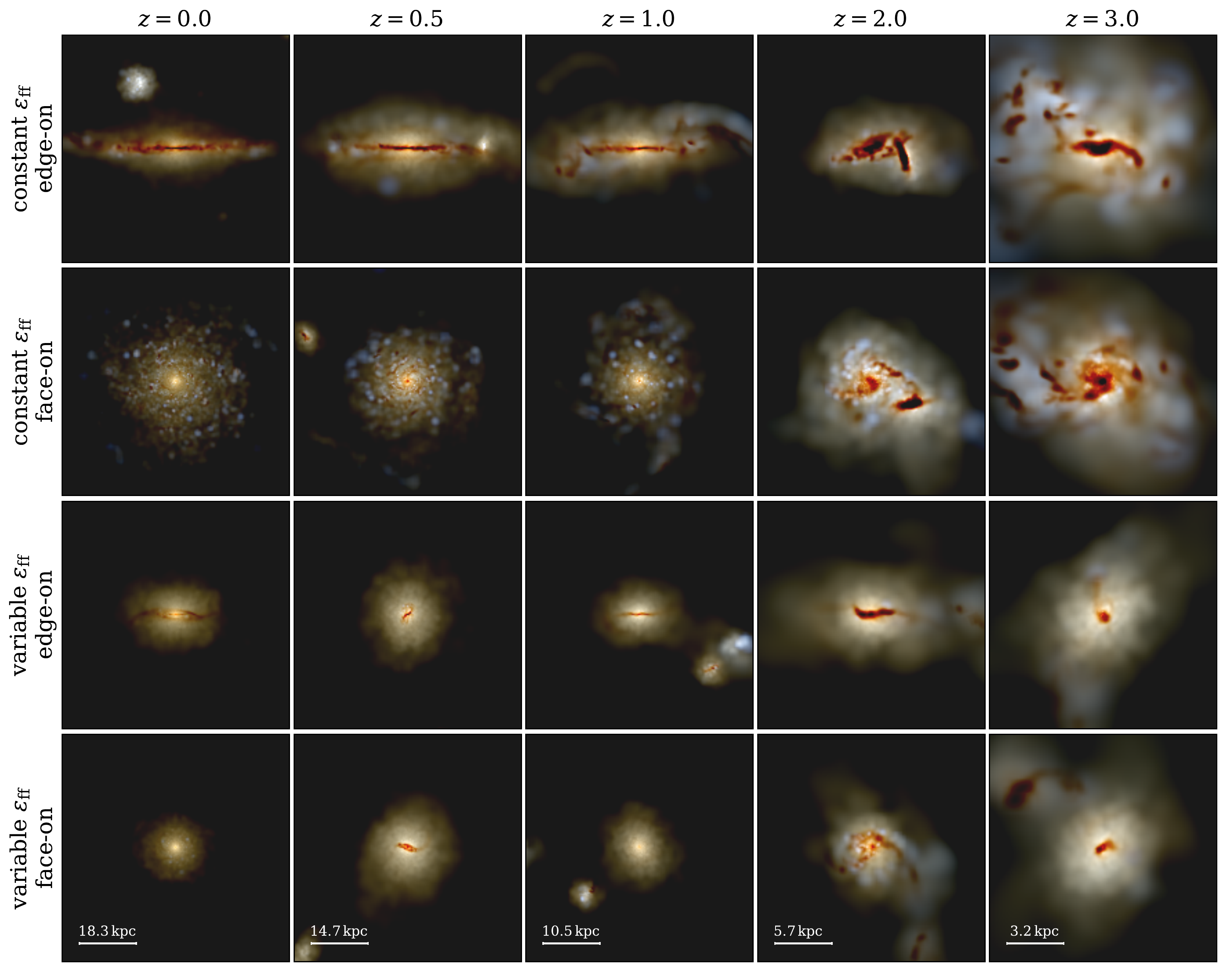}
\caption{The morphological evolution of the corresponding central galaxy using true colour images with dust absorption allowing us to render both stars and gas. Each column from left to right shows the galaxy at a different redshift ($z=0.0,0.5,1.0,2.0,3.0$). Top 2 rows show the evolution of the galaxy with the constant $\epsilon_\mathrm{ff}$ star formation recipe, edge-on and face-on. The result is the formation of a large gaseous and star forming disk.
Bottom 2 rows shows the evolution of the galaxy with the variable $\epsilon_\mathrm{ff}$ star formation recipe, edge-on and face-on. 
The resulting galaxy is a spheroidal galaxy with a tiny nuclear gas disk.}
\label{fig:render}
\end{figure*}

This effect can be partly explained by the higher gas and baryon fraction in the constant efficiency case.
Indeed, we see in the top panels of Figure~\ref{fig:render} the morphological evolution of the corresponding central galaxy using true colour images with dust absorption
that allows us to render both stars and gas. The constant efficiency case, owing to its higher gas and baryon fraction, leads to the formation of a large gaseous
and star forming disk. This is a well-known outcome of gas-rich mergers \citep{2008ApJS..175..356H,2010ApJ...713..686D}.
The varying efficiency case, on the other hand, leads to the formation of a spheroidal galaxy with a tiny nuclear gas disk (see bottom panels of Fig.~\ref{fig:render}). 
In this case, the gas fraction is much lower, leading to the formation of dispersion dominated systems as explained by the gas-poor, dry merger scenario 
\citep{2007ApJ...658..710N}.

The origin of this difference can also be found in the star formation efficiency itself. 
In the lower panel of \autoref{fig:sfr_eps}, we plot the mean efficiency of the star forming gas
within $0.1 \times R_{\rm vir}$, defined as 
\begin{equation}
\left< \epsilon_{\rm ff} \right> = \int \epsilon_{\rm ff} \frac{\rho}{t_{\rm ff}}{\rm d}V / \int\frac{\rho}{t_{\rm ff}}{\rm d}V 
\end{equation}
The constant efficiency with $\epsilon_{\rm ff}=0.01$ is also shown for comparison. We clearly see that during the starbursts the average
efficiency peaks at up to 3\%, while in the post-starburst quiescent phases, it drops down to 0.1\%. The high efficiency we obtain during 
the mergers explains how the galaxy, although relatively gas poor compared to the other model, manages to have a SFR as high as 10~$M_\odot~{\rm yr}^{-1}$,
like in the constant efficiency case. This helps maintaining the gas and baryon fractions low. 

Note that this high efficiency is significantly higher than the critical value
of 1\% proposed by \cite{2018ApJ...861....4S} that divides the feedback-dominated star formation regime from the efficiency-dominated star formation regime.
Our constant efficiency model with $\epsilon_{\rm ff}=0.01$ sits just at the limit, so the resulting galaxy never reaches the feedback-dominated regime.
Our varying efficiency model, on the other hand, clearly explores deep into these two regimes: 
1- strong feedback-dominated starbursts, for which the exact value of $\epsilon_{\rm ff}$
probably plays a minor role, as long as it is larger than 1\% and 2- long, extended, post-starbursts quiescent phases, for which the galaxy is quenched and the efficiency drops down to 0.1\%.

\begin{figure}
\includegraphics[width=\columnwidth]{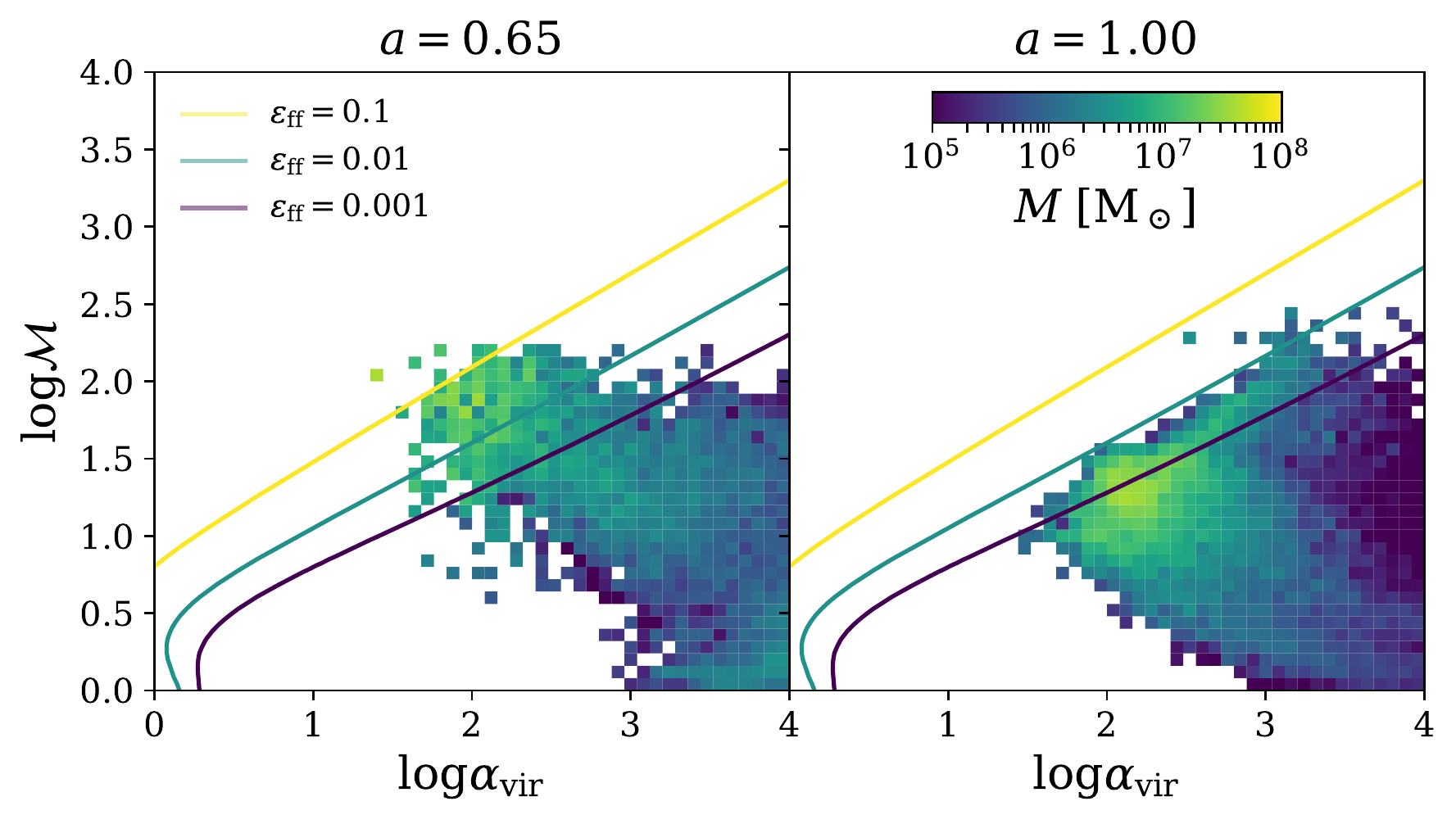}
\caption{Mass-weighted histogram in the $\alpha_\mathrm{vir}$ and $\mathcal{M}$ phase-space at two different epochs. Left:
the epoch of the starburst triggered by the last major merger. Right: the final epoch corresponding to the quenched, early-type galaxy.
Coloured lines correspond to constant star formation efficiencies.}
\label{fig:phase}
\end{figure}

In order to understand the origin of these wide variations in efficiency, we show in Figure~\ref{fig:phase} a mass-weighted histogram of the star forming gas
within $0.1 \times R_{\rm vir}$ in the $\alpha_{\rm vir}$-$\mathcal{M}$ plane. In the starburst case, at $a=0.65$, most of the star forming gas lies close
to the 10\% efficiency line. This is a combination of two factors: First, the merger event triggers the fast migration of gas to much higher density and
smaller $\alpha_{\rm vir}$ \citep[see e.g.][]{2010ApJ...720L.149T}. The mean gas density during the starburst is roughly $n_{\rm H} \simeq 20$~H/cc. 
In the same time, the merger favours a higher gas turbulent velocity dispersion, with a mass-weighted average value as high as $\sigma_{\rm 1D} \simeq 70$~km/s. 
Both effects work in tandem to maintain the efficiency around 10\% during the merger. This leads to extremely small depletion times, around 100~Myr,
but the gas is also quickly collapsing and this high efficiency state can be maintained during the duration of the merger.

In the quiescent phase, we see in Figure~\ref{fig:phase} that most of the gas sits close to the 0.1\% efficiency line. We are in the exact opposite situation than
the starburst: residual gas accretion maintains a non-negligible, albeit lower, level of turbulence, around 20~km/s, while in the same time the mean gas density dropped to
a rather low value with $n_{\rm H} \simeq 0.5$~H/cc. This leads to extremely long depletion times, around 10~Gyr. This residual star formation occurs mostly
in a small nuclear disk a few kpc in size. 

Milky Way-like galaxies correspond to the intermediate regime between the starburst and the quenched galaxy. We have explored this regime
in a companion paper using a different halo (Kretschmer {\it et al}. in prep) and indeed, the mean efficiency settles naturally to 1\% in the disk. Our varying efficiency model 
automatically adjusts its value to the conditions within the galaxy, each main galaxy type (starburst, disk, spheroid) giving rise to a different regime of star formation. 
This is due to a different structure of the ISM, with more or less gas being able to reach the limit of collapsing transonic cores.

\subsection{Effect of the subgrid turbulence model}
The last important ingredient in our subgrid galaxy formation model is the SGS turbulence model.
As explained in the previous section, we use a source term for the turbulent kinetic energy based on the local shear tensor
and a sink term based on dissipation of kinetic energy over one turbulent crossing time. 
In order to test the robustness of our star formation model to the adopted model for turbulence, we decided to change the shear-based source term 
classically used in the SGS model by a supernova-based source term, in the spirit of the work of \cite{2016ApJ...826..200S}. 
Note that in the classical SGS model, supernovae explosions will also inject turbulence indirectly though momentum deposition followed by the corresponding shear-induced source term. We nevertheless turn the shear source term off and replace it by direct turbulent kinetic energy 
being deposited in the cell where the supernova explodes, assuming 10\% of the energy in turbulent form. 

\begin{figure}
\includegraphics[width=\columnwidth]{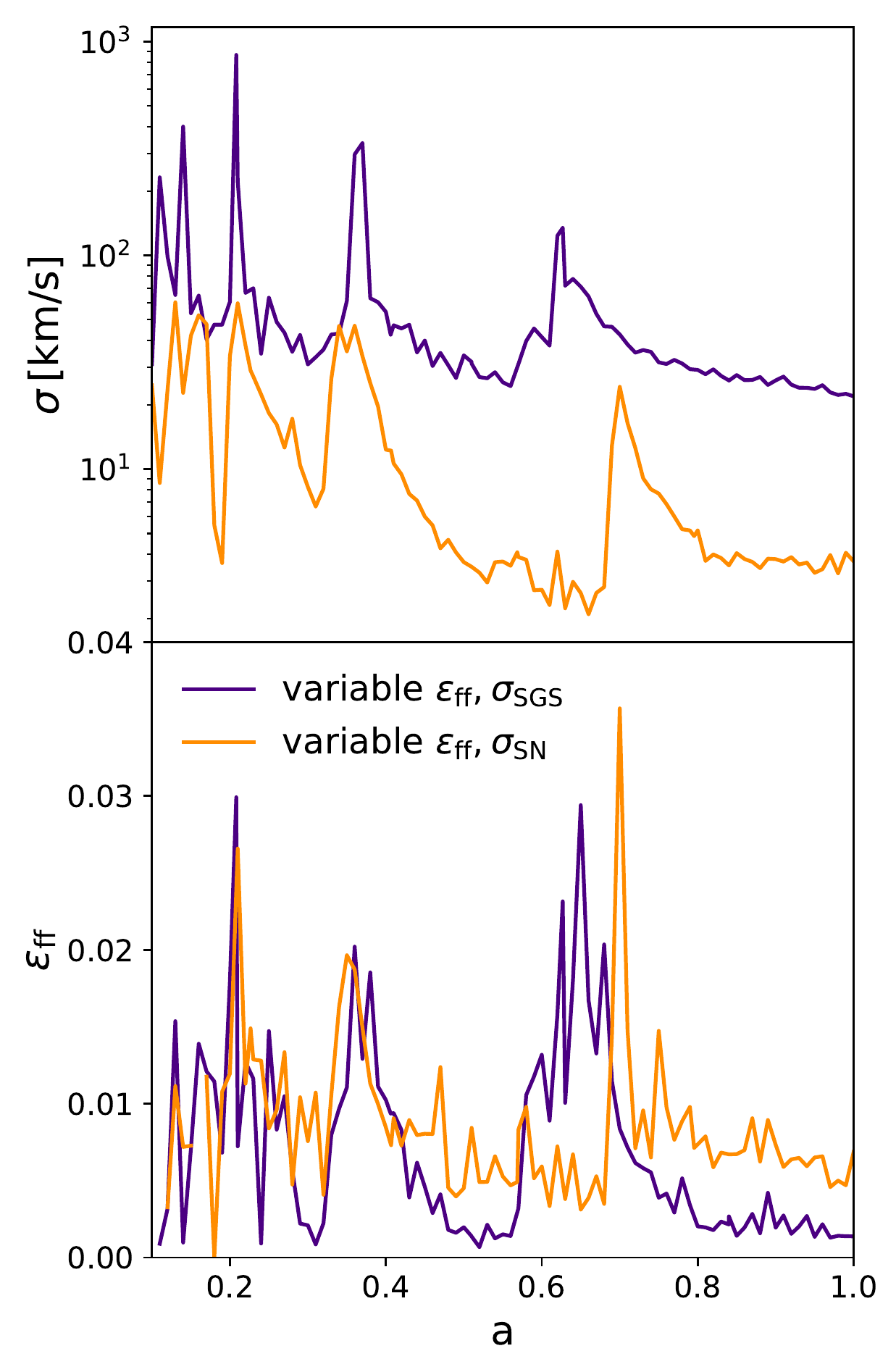}
\caption{Top panel: Average turbulent velocity dispersion in the ISM as function of scale factor for two different source terms for turbulence. 
The SGS source term is based on the local shear and gives rise to larger velocity dispersions, 
compared to a source term based on direct kinetic energy injection by supernovae (SN). 
Bottom panel: The average star formation efficiency as a function of scale factor for the two different turbulent source terms.}
\label{fig:turb_model}
\end{figure}

The top panel of \autoref{fig:turb_model} shows the evolution of the mass-weighted average turbulence resulting from these two models. 
The direct injection of turbulent energy from SN and the removal of the shear-based source term produces a level of turbulence 
roughly one order of magnitude smaller than the SGS model. 
Interestingly, this has a weak effect on the resulting star formation efficiency, as shown in the bottom panel of \autoref{fig:turb_model}. 
It is apparent that the difference between the two models is quite small, even though the actual level of turbulence is very different. 
This is because the varying efficiency model adapts to the local conditions in the galaxy and will transfer gas to smaller $\alpha_{\rm vir}$ (or higher density) 
compared to the SGS model, maintaining a similar star formation efficiency.

There is a small difference in scalefactor for the starburst at $a\simeq0.65$ which originates from slightly different trajectories caused by numerical effects \citep{2019MNRAS.482.2244K,2019ApJ...871...21G}.
Furthermore, the efficiencies for the SN induced turbulence model in the quiescent phases is larger compared to the efficiencies in the SGS run.
Consider for example the left panel of \autoref{fig:phase}. In the SN turbulent case, many cells will have much smaller $\alpha_{\rm vir}$ and $\mathcal{M}$, bringing the individual gas cells into a regime where star formation is dominated by the collapse criterion through $\alpha_{\rm vir}$. Therefore, during a quite phase, in the SN injected turbulence case the efficiencies are larger because the small values for the turbulence will bring cells into a regime where star formation is controlled by the virial collapse.

\begin{figure}
\includegraphics[width=\columnwidth]{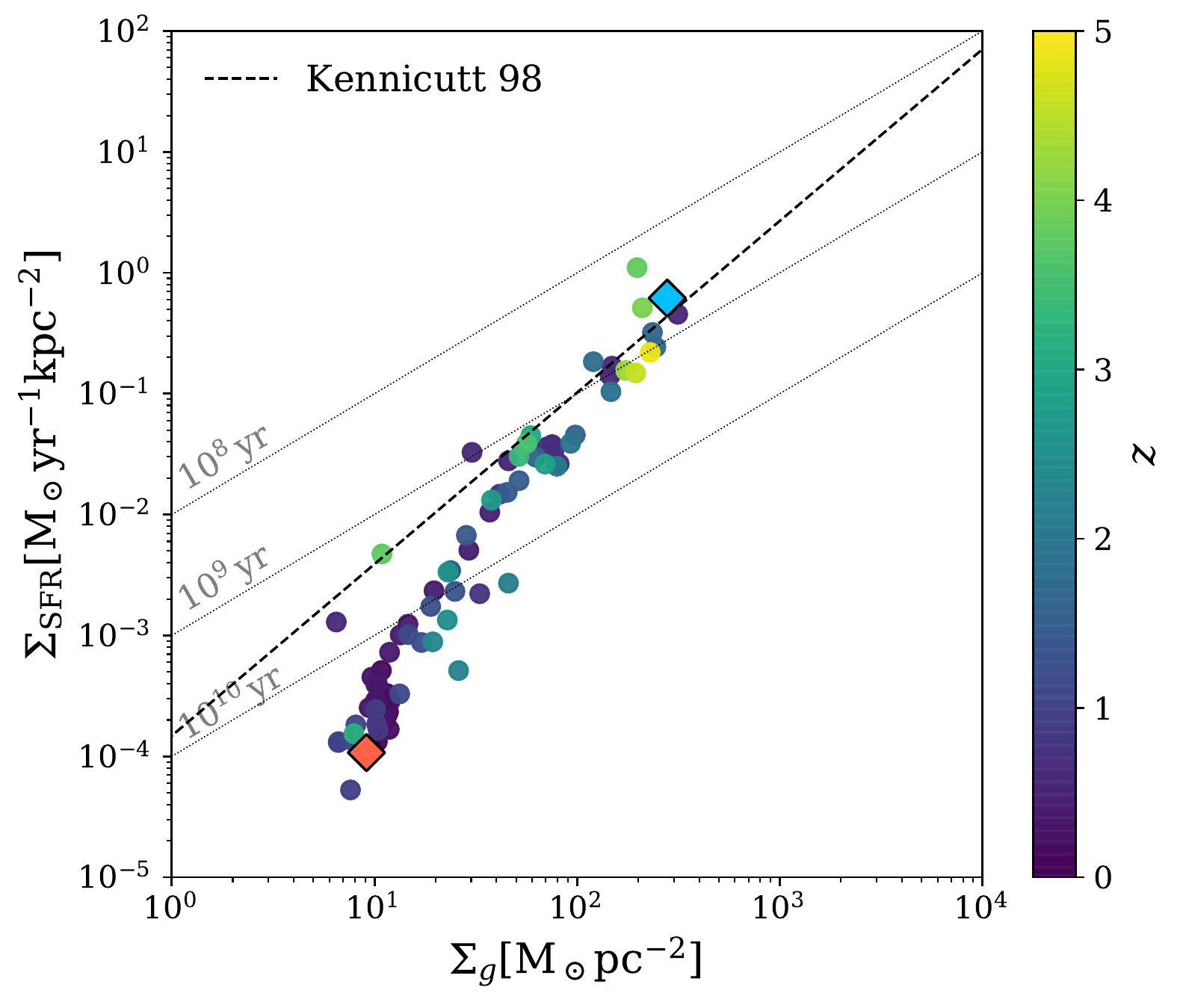}
\caption{The unresolved KS diagram in the simulations. Each point is the average surface star formation rate and surface densities corresponding to a snapshot where the colour indicates the redshift. (See the text for the exact definition). The thin lines indicate constant depletion times of $10^8, 10^9$ and $10^{10}$ years. The dashed line is the empirical KS-law from \citep{1998ApJ...498..541K}. The blue diamond symbol highlights the measured point during a starburst event at $a=0.65$ and the red highlights the obtained point for the quiescent phase $a=1.0$.}
\label{fig:globalKS}
\end{figure}

\begin{figure}
\includegraphics[width=\columnwidth]{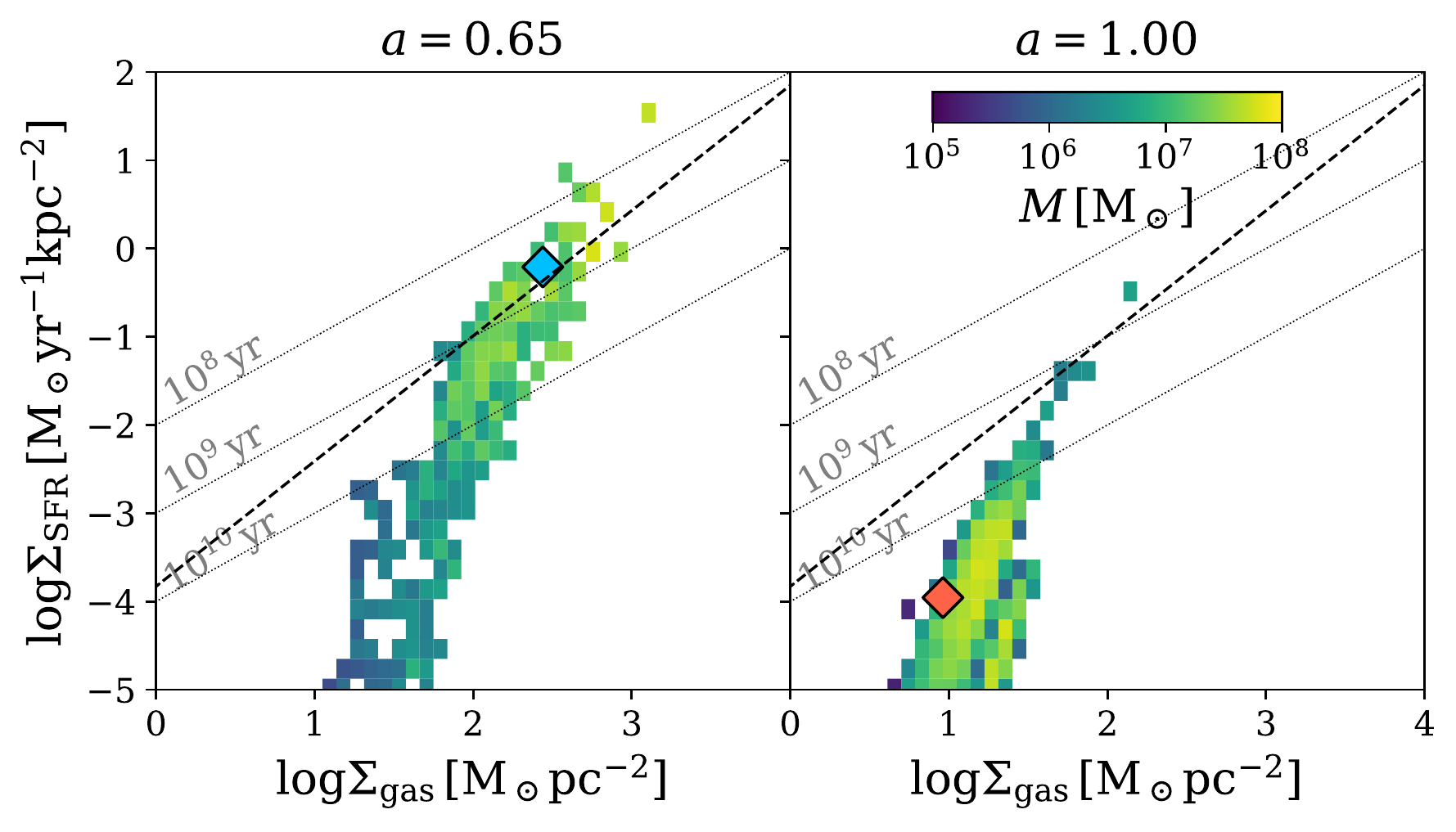}
\caption{Mass-weighted distribution of the gas mass in a resolved KS diagram. Star formation and gas surface densities are computed in square patches of 200pc. The thin lines indicate constant depletion times of $10^8, 10^9$ and $10^{10}$ years. The dashed line is the empirical KS-law from \citep{1998ApJ...498..541K}. Left is the distribution at $a=0.65$ during a starburst and right is the distribution at $a=1.0$ in a quiescent state.
The diamond symbols are the averaged (global) values and the same as in \autoref{fig:globalKS}.}
\label{fig:resolvedKS}
\end{figure}

%% file: discussion.tex
\section{Discussion}

In this paper, we have introduced a new set of subgrid models for star formation and feedback, uniquely combined in this new version of RAMSES.
The feedback model, very similar qualitatively to earlier implementations of terminal momentum injection by supernovae, but quite different in the details, provides
the first order mechanism for regulating the stellar and baryonic content in our simulated galaxies. The multi-freefall star formation model, based on a varying 
efficiency based on a subgrid turbulence model, allows us to form star without invoking arbitrary values for both the efficiency and the density threshold. 
We have shown that in a starburst regime, the efficiency can be significantly higher than 1\%, at odd with traditional models of galaxy formation, while in a
quenched regime, the efficiency can be as low as 0.1\%, owing to the very different nature of the turbulent ISM in post-starburst, early-type galaxies.

Our star formation model, based on an SGS model for unresolved turbulence, and the subgrid model for supernovae momentum feedback 
are naturally designed in such a way that they are (relatively) resolution independent. When resolution is increased, the kinetic energy of turbulence
will decrease, following the scaling for Burger's (or Larson's) turbulence. Ultimately, when the sonic length is resolved, the SGS turbulence will become
subsonic and our star formation recipe with turn into a simple collapse criterion for thermally supported gas with 100\% efficiency. 
At these scales, however, one usually adopts a sink particle formalism to describe star formation.

For supernovae feedback, increasing the resolution will allow us to resolve more the cooling radius, assuming the supernovae explode in a fixed density environment.
This won't be the case, unfortunately, as simulations with better and better resolution lead to denser and denser environments.
In the limit of sub-parsec resolution, we know from dedicated ISM studies \citep{2014ApJ...788..121K,2015MNRAS.454..238W,2017A&A...604A..70I} that a crucial ingredient is the inclusion of 
walk-away and run-away massive stars.
Overall, the galaxy formation simulations we present in this paper are far from resolving these scales, so that our subgrid models are crucial in providing a realistic
treatment of these unresolved phenomena. 

As explained in the previous section, our simulated halo allowed us to study two different regimes of star formation: a merger-induced starburst and a quenched early-type galaxy.
The SFR in these two extreme cases are widely different, with 10~$M_\odot~{\rm yr}^{-1}$ for the starburst and 0.1~$M_\odot~{\rm yr}^{-1}$ for the early-type. 
These variations can be explained partly by the different gas fraction, but also by the different typical density and Mach number in these galaxies. 
We now compare the properties of our simulated galaxy to observed ones using the global KS relation in Figure~\ref{fig:globalKS}.
For this, we compute the total star formation rate in the galaxy by integrating the star formation rate in each gas cell within $0.1 \times R_{\rm vir}$
and the total gas mass within the same region. We then use the half-mass radius of the gas $r_h$ to compute the global surface densities of both 
star formation and gas as \citep{2019A&A...622A.105F}
\begin{equation}
\dot \Sigma_* = 0.5 \frac{\dot M_*}{\pi r^{2}_h}~~~{\rm and}~~~\Sigma_{\rm gas} = 0.5 \frac{M_{\rm gas}}{\pi r^{2}_h}.
\end{equation}
In Figure~\ref{fig:globalKS}, each symbol corresponds to the location of the galaxy in the global KS diagram at a different epoch. 
We also indicate 3 lines corresponding to a constant depletion time of 100~Myr, 1~Gyr and 10~Gyr. 
Although high redshift conditions lead in general to small depletion times, the galaxy is moving quite widely along the KS relation \citep{2016MNRAS.457.2790T, 2018ApJ...859...56T}.
The two diamond symbols show the most extreme conditions we have obtained, with the starburst at the rightmost tip of the KS relation,
and the quenched early-type at the leftmost end. 

To understand better the origin of these two extreme cases, we show in Figure~\ref{fig:resolvedKS} the distribution of the gas mass
in a resolved KS diagram. For this, we compute the star formation and gas surface densities using square patches of 200~pc size
and compare it to the observed KS relation as well as constant depletion time models. In each case (starburst and early-type) we also show
as a single diamond symbol the global values. We clearly see that the gas and star formation surface densities is the starburst case are both very large,
with very small depletion times, around 1~Gyr in average but as small as 100~Myr in some regions. 
The early-type galaxy at the final epoch, on the other hand, has an average depletion time around 20~Gyr, with some regions reaching barely 1~Gyr.

In order to compare to molecular observations, we use the model of \cite{2018MNRAS.473..271V} to compute the molecular $\rm H_2$ mass within 
each cell, assuming for the interstellar radiation field a global uniform value given by $G=G_0 (\mathrm{SFR / M}_\odot {\rm yr}^{-1})$ at each epoch.
For the merger-induced starburst at $a \simeq 0.65$, we find $M_{\rm gas} \simeq 2 \times 10^9~M_\odot$ and $M_{\mathrm H_2} \simeq 9 \times 10^8~\rm M_\odot$,
while for the early-type quenched galaxy at $a=1$ we have $M_{\rm gas} \simeq 4 \times 10^9~\rm M_\odot$ and $M_{\rm H_2} \simeq 8 \times 10^8~\rm M_\odot$.
Note that our model for $\rm H_2$ formation is based on exactly the same turbulence-based subgrid model for star formation, so that there is a built-in correlation
between star formation and molecular gas formation, but no direct causal relation.

In the early-type case, the long depletion time coincides with a low $H_2$ fraction, so that the molecular gas depletion time becomes shorter by a factor of $\sim5$, around 20~Gyr. 
This is in perfect agreement with the COLD GASS sample observed at IRAM \citep{2011MNRAS.415...32S,2011MNRAS.415...61S}, with our early-type case corresponding to a low specific star formation rate $\mathrm{sSFR} \simeq 10^{-12}~{\rm yr}^{-1}$ and long molecular depletion times $t_{\rm dep}(\rm H_2) \simeq 20$~Gyr, and our starburst case corresponding to LIRGs or ULIRGs \citep[see Fig.~9 in][]{2011MNRAS.415...61S}. 

Interestingly, post-starburst galaxies show many similarities with our quenched mode of star formation. 
In several recent papers, galaxies showing signs of recent starburst activity are completely quenched,
although they contain enough gas to form stars at a rate much higher than observed \citep{2018MNRAS.476..122V,2018MNRAS.478.3447E,2018ApJ...855...51S}.
Residual turbulence in the ISM or high levels of radiation were proposed as possible explanations to quench star formation in such systems. 
These systems are probably ideal examples of galaxies with a very peculiar ISM structure, leading to an exceptionally low star formation efficiency.

At late times the baryon fraction in the halo stays constant. At the same time the gas fraction in the disk increases by $\sim 10\%$ due to gas cooling from the halo onto the disk. The star formation rate however does not increase with the gas fraction but rather decreases. This is because the efficiency depends on the shear-induced turbulence in a non-linear way. Namely the level of turbulence in the disk from accretion or shear will prevent star formation.
In contrast, the star formation rate given by the constant efficiency model does increase with the gas fraction of that run.

Another interesting class of objects are circumnuclear disks at the centre of early-type galaxies. Our simulation with varying efficiency 
shows at the final epoch such a nuclear disk (see Fig.~\ref{fig:render}). Nuclear disks in early-type galaxies are compact, relatively dense in their centre 
with $\Sigma_{\rm gas} \simeq 100~\rm M_\odot~{\rm pc}^{-2}$ but very inefficient in forming stars \citep{2014MNRAS.444.3427D},
when compared to normal galaxies with the same gas surface densities. The strong shear observed in these systems was proposed as 
an explanation for their surprisingly low star formation efficiency. One particularly striking example is NGC~4429 from the same ATLAS3D sample:
In a recent attempt to model it, Naab and co-workers (private communication) had to use a constant efficiency of $\epsilon_{\rm ff} \simeq 0.2\%$
to reproduce various properties such as gas content and star formation rate. This is very close to the efficiency we obtained using the 
varying efficiency model in our final early-type galaxy. 

%% file: conclusion.tex
\section{Summary and Conclusion}
In this paper, we have introduced a new implementation of recently proposed subgrid models for star formation and feedback in the RAMSES code. 
The stellar feedback model is based on individual, discrete Type II supernova events. If the radius marking the transition from the energy-conserving phase 
to the momentum-conserving phase is unresolved by our computational cell, we inject directly momentum in the neighbourhood of the exploding star. 
This is done by modifying the Riemann solver in our Godunov scheme, allowing us to deliver the right amount of momentum, together with consistent mass and energy fluxes.

The multi-freefall star formation model is based on a subgrid turbulence model and provides a varying star formation efficiency that depends on the local conditions in each 
computational cell, through the cell's virial parameter and its turbulent Mach number. 
The star formation efficiency can be very large for large density fluctuations caused by supersonic turbulence, as expected, for example, during a merger. 
On the other hand, the star formation efficiency can be very low for subsonic and marginally stable disks, as expected during more quiescent phases or in early-type galaxies. 
This new versatile model allows us to form star without invoking arbitrary values for both the efficiency and the density threshold.

We apply this new implementation of subgrid galaxy formation physics to a prototype cosmological simulation of a massive halo that features a major merger, 
and study specifically the interesting case of the formation of an early-type galaxy.
We find that the feedback model provides the first order mechanism for regulating the stellar and baryonic content in our simulated galaxies.
Together with the multi-freefall star formation model, feedback is strong during starbursts. At redshift zero, only 40\% of baryons are retained in the halo and the gas fraction is $\sim 20\%$. 
The resulting stellar masses are in good agreement with those expected from abundance matching results.

In a starburst regime, the efficiencies can be significantly higher than 1\%, at odd with traditional models of galaxy formation, while in a quenched regime, the efficiencies can be as low as 0.1\%, 
owing to the very different nature of the turbulent ISM in post-starburst, early-type galaxies.
Indeed, the merger driven event pushes gas to large densities and large turbulent velocity dispersions, which causes larger $\mathcal{M}$ and small $\alpha_\mathrm{vir}$. 
As a result, efficiencies can reach locally $10\%$ and the $\mathrm{sSFR}$ is large. 
The small value for the molecular gas depletion time during the starburst is in perfect agreement with observations. 
Additionally, at late times, when the galaxy is quiescent, the $\mathrm{sSFR}$ is small and the molecular gas depletion time is long, also in perfect agreement with observations.

In summary, at high redshifts, very efficient star formation together with strong feedback regulates the baryonic content. 
At low redshift, local star formation becomes very inefficient explaining the properties of quenched galaxies.

%% file: acknowledgements.tex
\section*{Acknowledgements}
The authors thank the referee for their constructive comments that improved the quality of the paper.
We acknowledge stimulating discussions with Robert Feldmann, Lucio Mayer, Robbert Verbeke, Renyue Cen and Sandro Tacchella. 
This work was supported by the Swiss National Supercomputing Center (CSCS) project s890 - ``Predictive models for galaxy formation'' 
and the Swiss National Science Foundation (SNSF) project 172535 - ``Multi-scale multi-physics models of galaxy formation''.
The simulations in this work were performed on Piz Daint at the Swiss Supercomputing Center (CSCS) in Lugano,
and the analysis was performed with equipment maintained by the Service and Support for Science IT, University of Zurich.
We also made use of the pynbody package \citep{2013ascl.soft05002P}.